\documentclass[reprint, amsmath, amssymb, aps, pra, superscriptaddress]{revtex4-2}

\usepackage[colorlinks=true]{hyperref}
\usepackage{bm}
\usepackage{physics}
\usepackage{subcaption}
\usepackage{graphicx}
\usepackage{soul}
\usepackage{comment}
\usepackage{ulem}
\usepackage{caption}   

\usepackage{adjustbox}

\usepackage{appendix}

\newcommand{\mcA}{\mathcal{A}}

\begin{document}

\preprint{APS/123-QED}

\title{Photon emission by vortex particles accelerated in a linac}

\author{A.\,Yu. Murtazin}
\email{anton.murtazin@metalab.ifmo.ru}
\affiliation{School of Physics and Engineering,
ITMO University, 197101, St. Petersburg, Russia}

\author{G.\,K. Sizykh}
\affiliation{School of Physics and Engineering,
ITMO University, 197101, St. Petersburg, Russia}
\affiliation{Petersburg Nuclear Physics Institute named by B.P. Konstantinov of National Research Centre ``Kurchatov Institute'', 188300, Gatchina, Russia}

\author{D.\,V. Grosman}
\affiliation{School of Physics and Engineering,
ITMO University, 197101, St. Petersburg, Russia}

\author{U.\,G. Rybak}
\affiliation{Peter the Great St. Petersburg Polytechnic University, 194021, St. Petersburg, Russia}

\author{A.\,A. Shchepkin}
\affiliation{School of Physics and Engineering,
ITMO University, 197101, St. Petersburg, Russia}
\affiliation{Petersburg Nuclear Physics Institute named by B.P. Konstantinov of National Research Centre ``Kurchatov Institute'', 188300, Gatchina, Russia}

\author{D.\,V. Karlovets}
\email{dmitry.karlovets@metalab.ifmo.ru}
\affiliation{School of Physics and Engineering,
ITMO University, 197101, St. Petersburg, Russia}
\affiliation{Petersburg Nuclear Physics Institute named by B.P. Konstantinov of National Research Centre ``Kurchatov Institute'', 188300, Gatchina, Russia}

\begin{abstract}
We study the photon emission by charged spinless particles with phase vortices and an orbital angular momentum (OAM) projection in longitudinal electric and magnetic fields within the scalar QED. A realistic wave packet of an electron or ion accelerated by a radio-frequency wave locally feels a constant and spatially homogeneous field, which allows us to develop an effective model for losing the angular momentum of the vortex particle due to photon emission. For the fields typical for accelerator facilities, we find that an effective lifetime of the vortex state greatly exceeds the acceleration time. This proves that the acceleration of vortex electrons, ions, muons, and so forth to relativistic energies is possible in conventional linacs, 
the OAM losses due to the photon emission are mostly negligible, and that the vortex quantum state is highly robust against these losses. 
\end{abstract}

\maketitle

\section{Introduction}

The emission of photons in quantum field theory (QFT) by charged particles accelerated in an external electromagnetic field is an inherently nonstationary process because acceleration occurs during a finite time interval over a finite space domain. In contrast to that, the customary relativistic perturbation theory within the S-matrix formalism deals with the stationary quantum states of particles, so the emission process formally occurs during the infinite time interval with no spatial localization. It is only when the particle path is periodic -- say, a circle in a magnetic field -- that the stationary perturbation theory successfully describes the emission \cite{SokolovTernov}. The difference between the stationary and nonstationary approaches can be illustrated with the formation length (zone) effects when photons are emitted by relativistic charged particles in media, such as, for instance, the Landau-Pomeranchuk-Migdal effect, pre-wave zone effect, etc. \cite{BAIER2005261, Polytitsyn_diffraction}. The use of the standard stationary theory hinders the analysis of the formation-length and time phenomena. 

Although the quantum states of charged particles accelerated in combinations of electric and magnetic fields have long been known \cite{BagrovGitman, FradkinGitman_unstable_vacuum}, their applications to specific calculations in QED are extremely scarce. Leaving aside the physical interpretation of such states, there is a fundamental difficulty in dealing with the accelerated particles. When the mean energy exceeds the double rest energy of an electron, $2mc^2$, the vacuum can become unstable and the particle-antiparticle pairs can start to create \cite{Nikishov1969_pair_production, FradkinGitman_unstable_vacuum}. To deal with the realistic accelerator facilities with fields much weaker than the Schwinger critical value $E \ll E_c = m^2c^3/|e|\hbar= 1.32\cdot10^{16}\,\text{V}/\text{cm}$, we need to employ 1-particle states with a stable vacuum and therefore, somewhat contrary to what the QED community has been doing during the last several decades \cite{FradkinGitman_unstable_vacuum} -- to get rid of the pairs.

Here, we develop {\it an effective model} of photon emission by scalar charged particles with phase vortices in relativistic nonstationary perturbation theory. In modern facilities, acceleration occurs in radio frequency cavities in which particles are pushed by electric impulses from traveling waves. The key feature that allows us to significantly simplify the model is that the wave packets of electrons, muons, protons, ions, etc. are {\it many orders of magnitude smaller} than {\it (i)} the length of an accelerating wave in a waveguide, {\it (ii)} spatial inhomogeneities of the accelerating fields. Therefore, if we suppose that the mean group velocity of the particle packet always matches the phase velocity of the accelerating wave, the particle {\it locally experiences a constant and homogeneous electric field} ${\bm E} = \{0,0,E\}$ directed along the acceleration axis $z$. 

The exact solutions to both the Klein-Gordon and Dirac equations in such a field are known, see \cite{BagrovGitman}. However, these solutions can be inconvenient in practical calculations due to admixture of the electron-positron pairs (for brevity, we call the particles electron and positron even if they have no spin). Instead, we develop an iterative procedure following the Wentzel–Kramers–Brillouin (WKB) or quasi-classical approach \cite{MaslovFedoruk,AkhiezerBerestetsky,Bagrov1993} that keeps us within a class of 1-particle states with stable vacuum. The corresponding quasi-classical state no longer obeys the Klein-Gordon equation exactly, but the accuracy stays under control and it is enough to analyze the emission of a photon within the first order of the nonstationary perturbation theory. 

In particular, the finite time interval during which acceleration occurs $t_f-t_i$ is embedded in this state. This {\it temporal localization} inevitably implies {\it spatial localization} of the wave packet in the longitudinal direction, i.e., the existence of the packet center defined by a mean classical path $\langle {\bm r}\rangle(t)$ \cite{Bagrov1993}, the latter being the relativistic hyperbolic trajectory in the electric field \cite{LandauLifshitz2}.

We employ these quasi-classical states of accelerated charged spinless particles with phase vortices, dubbed vortex or twisted particles \cite{Bliokh2017, IVANOV2022}, for studying photon emission within the first order of the perturbation theory in scalar QED. For the electric and magnetic field strengths typical for conventional accelerators, the time during which the vortex particle loses its orbital angular momentum (OAM) due to photon emission is found to be significantly greater than the typical time-of-flight in an accelerator, even for kilometer-long linacs. This proves the stability of the twisted quantum states of the charged particles against photon emission during the acceleration process and demonstrates the principal possibility to accelerate vortex electrons, muons, ions, etc. in a linac. Once this acceleration to relativistic energies is achieved, it opens new possibilities for studying a plethora of quantum phenomena beyond the reach of conventional plane-wave (or Gaussian) states without phase vortices \cite{Bliokh2017, IVANOV2022, Karlovets2021Vortex, KarlovetsSerbo2020, IvanovSuperkick2024, IvanovSuperkick2022, VortexMuon, Ivanov2020}.

Throughout this paper, we employ the system of units $\hbar = c = 1$, except in key formulas where retaining these constants is essential to construct the WKB solution.


\section{Loss of vorticity during acceleration}

Some of us have already argued that if a charged particle with a phase vortex is accelerated to relativistic energies, its quantum state with an orbital (or total) angular momentum (AM) projection $l$ stays very stable against the emission of twisted photons carrying the AM away \cite{Zmaga2025, Karlovets2021Vortex, DKGrosmanPavlovArxiv2025}. One can start with simple considerations from relativistic kinematics: in a longitudinal magnetic field of a focusing solenoid, the state of an electron with an AM $l$ lives a finite time interval (a lifetime) and decays due to emission of a twisted photon. If the total angular momentum of that photon $l_{\gamma}$ is maximized, $l_{\gamma}=l$, the electron goes to the final ground state with $l_f=0$ and thereby loses its vorticity. Suppose that we know the time $\tau$ during which this happens for a non-relativistic electron (see the corresponding calculations in \cite{Karlovets2023, Pavlov2024}). Then the lifetime in a laboratory frame in which this particle moves with a Lorentz factor $\gamma = \varepsilon/mc^2$ is
\begin{equation}
    t = \gamma\,\tau > \tau,
\end{equation}
i.e., the higher the energy, the less probable the emission becomes and the more stable the quantum state of the electron against the emission of twisted photons. 

The acceleration can be taken into account by integrating the emission
probability \textit{without electric field} over the
electron’s initial longitudinal momentum $p_z$, as done in
Ref.~\cite{DKGrosmanPavlovArxiv2025}. In other words, different momentum
components are combined \textit{incoherently}, at the level of probabilities.
In the present nonstationary approach the accelerated electron is described with
a single wave packet in the external field, and the $S$-matrix element
contains a time integral over the entire acceleration interval, see
Eq.~\eqref{eq:time_integrals}. When squaring the amplitude, one obtains terms
that represent interference between emission amplitudes associated with
different times along the accelerated trajectory. It is precisely this
temporal coherence that is absent in the probability-level treatment of Ref.~\cite{DKGrosmanPavlovArxiv2025}. In this paper we employ a more general theory in which this interference takes place.

We find that even with the above interference taken into account the
qualitative conclusions of Refs.~\cite{Karlovets2021Vortex, Karlovets2023,Pavlov2024,DKGrosmanPavlovArxiv2025, Meng2025PRR}
remain valid: the vortex states are extremely stable against OAM-changing
photon emission within the parameter range relevant for accelerator facilities. The probability of emitting a photon that changes the electron OAM during a single pass through the accelerator is very small, and the resulting
lifetime $\tau$ of the OAM state greatly exceeds the characteristic transit
time through the accelerating structure. 

Moreover, the hierarchy of channels
is robust: transitions with the small OAM loss $\Delta l = l-l' \sim 1$ dominate, whereas
channels with large $\Delta l \gg 1$ and those with change of the radial quantum number $\Delta n\neq 0$ are strongly
suppressed. At the same time, our approach reveals several important features
absent in simpler kinematic arguments. First, the emission probability and the
lifetime exhibit a significant dependence on the external electric field and on
the coherence length of the wave packet. For realistically short packets, with lengths on the order of $\sigma \sim 1-10 ~\text{nm}$ \cite{Ehberger, Cho2013, Karlovets2021Vortex}, the probability is practically independent of the electric field, whereas
longer packets acquire a narrow-region dependence on the field strength. Even in
the presence of very strong electric fields, the qualitative picture remains
unchanged: in all regimes we have explored, the probability of OAM-changing
emission stays tiny, and the corresponding lifetimes remain orders of
magnitude larger than the characteristic acceleration time.

In the recent paper by Meng \textit{et al.}~\cite{Meng2025PRR}, the
robustness of vortex states in a homogeneous electric field has also been emphasized. 
In the paraxial regime, the authors obtained an exact expression for dynamics of a beam width in electric field, including its dependence on
the field strength, initial momentum, and the initial width. The electric field has been shown to suppress free-space packet spreading, which can also be seen as a hallmark of the robustness of the vortex state during acceleration. A complementary analysis has also been recently done not only for electric field, but also for a more realistic RF accelerating wave in a linac \cite{Dyatlov2025}. In contrast to Ref.\cite{Meng2025PRR}, here we study the radiative dynamics of the vortex electron packet in a combination of electric and magnetic fields, taking quantum photon emission into account with the change of the electron OAM, as well as the packet spreading and its deceleration at high energies.




\section{Model}

\subsection{Quantum state of a charged particle in longitudinal electric and magnetic fields}

\subsubsection{Klein-Gordon equation with longitudinal fields}

We model electron dynamics in a linac via the Klein-Gordon equation in the longitudinal electric and magnetic fields:
\begin{equation}
\label{eq:KG_general}
    \left[ \left(i\partial^{\mu} - e \mcA^{\mu} \right)^2 - m^2 \right] \Psi(\bm{r},t) = 0.
\end{equation}
The electron is accelerated by the electric field along the $z$ axis, ${\bm E}=\{0,0,E\}$, and the magnetic field ${\bm H}=\{0,0,H\}$ is used to control the wave packet spreading in the transverse directions, similarly to a focusing magnetic lens. The field direction is chosen such that the particle accelerates along the $z$ axis; for an electron this corresponds to $E=-|E|<0$. The vector potential for stationary and homogeneous collinear electric and magnetic fields can be chosen in a Coulomb gauge as following
\begin{equation}
\label{eq:Potential_general}
    \mcA^{\mu} = \left\{0, \frac{H \rho}{2} \bm{e}_{\phi_r} + E t \bm{e}_z\right\},
\end{equation}
where $\bm{e}_z = \{0,0,1\}$ and $\bm{e}_{\phi_r}= \{-\sin\phi_r, \cos \phi_r, 0  \}$ are the basis vector along the $z$-axis and the azimuthal direction, respectively. Alternatively, in a stationary gauge
\begin{equation}
    \mcA^\mu = \left\{Ez, \frac{H\rho}{2}\bm{e}_{\phi_r}\right\},
\end{equation}
which provides the same field configuration. 
The results obtained below agree for both gauges up to a transformation that induces a phase shift in the wave function (App. \ref{app:classical_action}). 
For definiteness, let us take the gauge \eqref{eq:Potential_general}.
\newpage

Substituting \eqref{eq:Potential_general} into Eq. \eqref{eq:KG_general} leads to
\begin{widetext}
\begin{equation}
\label{eq:KG_expanded}
    \left( -\partial^2_t + \Delta - m^2 - \frac{e^2 H^2 \rho^2}{4} - e^2 E^2 t^2 -i e H\partial_{\phi_r} - 2 i e E t \partial_{z} \right) \Psi(\bm{r},t) = 0.
\end{equation}
\end{widetext}
Equation \eqref{eq:KG_expanded} admits the separation of variables, so the wave function is factorized:
\begin{equation}
\label{eq:state_general}
    \Psi(\bm{r}, t) = \Psi_{\parallel}(z, t) \Psi_{\perp}(\rho, \phi_r).
\end{equation}
Substituting the factorized form of the solution \eqref{eq:state_general} into Eq. \eqref{eq:KG_expanded},
we obtain two independent equations: one for the transverse mode $\Psi_\perp(\rho,\phi_r)$
and one for the longitudinal dynamics $\Psi_\parallel(z,t)$. Importantly, the transverse dynamics {\it depends only on the magnetic field} and not on the electric one, whereas for the longitudinal motion it is vice versa.

\subsubsection{Transverse dynamics}

Let us first examine the transverse dynamics, which is described by the solution $\Psi_{\perp}(\rho, \phi_r)$ to the equation
\begin{equation}
\label{eq:KG_transverse}
    \left(\Delta_\perp - \frac{e^2 H^2 \rho^2}{4} + e H \hat{l}_z + C\right) \Psi_{\perp}(\rho, \phi_r) = 0,
\end{equation}
that follows from Eq.(\ref{eq:KG_expanded}). Here, $\hat{l}_z = -i \partial_{\phi_r}$ is the operator for the z-projection of angular momentum, $C$ is the separation constant, which will be defined later. 
The exact solutions to Eq.~\eqref{eq:KG_transverse} with a well-defined orbital angular momentum (OAM) are the
\textit{scalar Landau states}:
\begin{equation}
\label{eq:state_transverse}
    \begin{aligned}
        \Psi_{\perp}(\rho, \phi_r) 
        &\equiv \Psi_{n,l}(\rho,\phi_r) \\
        &= N_{\perp}
        \left( \frac{\rho}{\rho_{\text{H}}} \right)^{l}
        L_n^l \!\left( \frac{2 \rho^2}{\rho_{\text{H}}^2} \right)
        \exp{-\frac{\rho^2}{\rho_{\text{H}}^2} + i l \phi_r},
    \end{aligned}
\end{equation}
where $N_{\perp}$ is the normalization constant (see App.~\ref{app:transverse_norm}),
and $\rho_{\text{H}}$ is the characteristic transverse scale (magnetic length) that 
sets the rms-radius of the ground state. It is given by
\begin{equation}
\label{eq:magnetic_length}
    \rho_{\text{H}} = \sqrt{\frac{4}{|e H|}} \equiv2 \lambda_c\sqrt{\frac{H_c}{H}}.
\end{equation}
For convenience, the critical magnetic field $H_c = m^2/|e| = 4.41 \cdot 10^9$ T and the electron Compton wavelength $\lambda_c = 1/m = 3.8 \cdot 10^{-11}$ cm are introduced.
The quantum numbers $n$ and $l$ are the radial and orbital indices,
respectively: $n = 0,1,2,\dots$, $l = 0,\pm 1,\pm 2,\dots$.

This solution dictates an exact form for the constant $C$:
\begin{equation}
    C = \frac{8}{\rho^2_\text{H}}\left(n + l + \frac{1}{2}\right),
\end{equation}
Additionally, introducing the transverse kinetic momentum operator, it can be shown that \cite{BagrovGitman}(App. \ref{app:transverse_momentum})

\begin{equation}
    \hat{\vb{p}}_\perp^2 \Psi_\perp(\rho,\phi_r) = C \Psi_\perp(\rho,\phi_r).
\end{equation}
This means $C \equiv p_\perp^2$ can be referred to as the conserved transverse momentum of a particle that can change only during spontaneous emission of photons, exactly as in the quantum theory of synchrotron radiation \cite{SokolovTernov}. Therefore, the transverse dynamics of this scalar charged particle is quantized (relativistic Landau states).

\subsubsection{Longitudinal dynamics}

As follows from Eq.(\ref{eq:KG_expanded}), the longitudinal wavefunction $\Psi_{\parallel}(z, t)$ is a solution to the following equation:
\begin{equation}
\label{eq:KG_longitudinal}
\left(-\partial_t^2 + \partial_z^2 - m^2 - e^2 E^2 t^2- 2 i e E t \partial_z - p_\perp^2 \right) \Psi_{\parallel}(z, t) = 0,
\end{equation}
where there is no magnetic field. This equation admits exact solutions in the form of parabolic cylinder functions $D_\nu(z)$ \cite{Nikishov1969, BagrovGitman}. However, as we have argued, they are inconvenient for our purposes. First, they are delocalized in space, and therefore describing acceleration of a charged particle with them during a finite time interval can be challenging even putting aside the fact that they contain an admixture of positrons. Second, it can be shown that the typical values for the index of the corresponding parabolic cylinder functions reach $\sim 10^6$ and even more for modern linacs. This makes the use of these states extremely inconvenient even for numerical simulations. 

To avoid this problem, we employ a semiclassical approach and solve Eq. \eqref{eq:KG_longitudinal} within the WKB-approximation, seeking the solution as a series in powers of $\hbar$ \cite{AkhiezerBerestetsky, Bagrov1993}. 
The effects of particle-antiparticle pair creation are hidden in the highest orders of this series, and we deliberately neglect them as they are negligible in the fields, typical for a realistic accelerator. The WKB approach allows us to construct a localized and normalized wave packet solution. To proceed, we recover the $\hbar$ dimension. The longitudinal part \eqref{eq:KG_longitudinal} of the Klein-Gordon equation \eqref{eq:KG_general} can be written in a more compact way as following:
\begin{equation}
    \left[\left(i\hbar\partial_\mu-\frac{e}{c}(\mcA_{||})_\mu\right)^2-m^2c^2 - p_\perp^2\right]\Psi_\parallel(z,t) = 0,\label{eq:longitudinal_wkb}
\end{equation}
 where $(A_{||})^\mu$ is a longitudinal part of the 4-potential vector: 
\begin{equation}
    (A_{||})^\mu = \{0,ctE\bm{e}_z\}.
\end{equation}

 According to the WKB approach, the solution can be found according to the following ansatz:
\begin{equation}
    \Psi_\parallel (z,t) = \left(\sum\limits_{k=0}^\infty\hbar^ka_k(z,t)\right) \exp{\frac{i}{\hbar}S_\text{cl}(z,t)},
\end{equation}
where $a_k$ are the coefficients to be defined, $S_\text{cl}$ is a classical action, solution of a Hamilton-Jacobi problem for a classical particle in a homogeneous electric field \cite{AkhiezerBerestetsky, Bagrov1993}. The longitudinal dynamics within the accuracy of $\mathcal{O}(\hbar^2)$ (see App. \ref{app:longitudinal_solution}) is:
\begin{equation}
\label{eq:state_longitudinal}
    \begin{aligned}
        & \Psi_{\parallel}(z, t) = \frac{N_{\parallel}}{\sqrt{\mathcal{E}_{\text{cl}}(t)}} \exp {\frac{i}{\hbar} S_{\text{cl}}(z, t) - \frac{(z - z_{\text{cl}}(t) )^2}{2 \sigma^2}}, \\
        & S_{\text{cl}}(z, t) = p_z z - \int\limits_{0}^t \mathcal{E}_{\text{cl}}(\tau) \dd \tau, \\
        & z_{\text{cl}}(t) = \frac{1}{F_0} \left( \mathcal{E}_{\text{cl}}(t) - \mathcal{E}_{\text{cl}}(0) \right),\\
        &\mathcal{E}_\text{cl}(t) = \sqrt{m^2c^4 + c^2p_\perp^2+ c^2(p_z+F_0t)^2}.
    \end{aligned}
\end{equation}
Here, $N_{\parallel}$ is a normalization constant (App. \ref{app:longinudinal_norm}), $\mathcal{E}_{\text{cl}}(t)$, $S_{\text{cl}}(z, t)$, $z_{\text{cl}}(t)$ are the classical energy, action, and the mean trajectory of a charged particle accelerating along the $z$-axis (cf. Ref.\cite{LandauLifshitz2}). Here,
\begin{equation}
    F_0 = -eE,
\end{equation} 
is a classical force acting on a particle. For an electron $F_0 = - |F_0| <0$. The parameter $\sigma$ is the coherence length (or simply length) of the electron packet along the z axis in the laboratory reference frame. We set the time instant for the particle entering the electric field region to be $t = 0$. The accuracy of the solution \eqref{eq:state_longitudinal} is analyzed in App. \ref{app:accuracy} in detail.

\subsection{Emission of a photon by a charged scalar particle}

\subsubsection{Matrix element in scalar QED}

In scalar QED, the first-order transition matrix element
for the emission of a plane-wave photon is
\begin{equation}
\label{eq:Sfi_general}
    \begin{aligned}
        \displaystyle S_{fi} = & -ie\int d^4x\, j_{fi}^{\mu}(x)A_{\mu}^*(x),\cr
        \displaystyle j_{fi}^{\mu} = \frac{1}{2m}\Psi_f^*(x) & (i\partial^{\mu} -e\mathcal A^{\mu})\Psi_i(x) + \\
        & + \frac{1}{2m}\Psi_i(x) (i\partial^{\mu} -e\mathcal A^{\mu})^*\Psi_f^*(x) = (j_{fi}^{\mu})^{\dagger}, \cr
        \displaystyle A_{\mu}^*(x) = & \frac{1}{\sqrt{2\omega V}}\, e_{\mu}^*(k)\,e^{ikx},\ e_{\mu}k^{\mu} = 0.
    \end{aligned}
\end{equation}
Here $j_{fi}^{\mu}$ is the transition current from the initial state $\Psi_i(x)$ to the final one $\Psi_f(x)$; $A_{\mu}$ is the 4-potential of the emitted photon with the wave vector $k$, frequency $\omega$ and the polarization vector $e^{\mu}$; $V$ is the normalization volume. The photon wave 4-vector is
\begin{equation}
     k^{\mu} = \omega\{1, \sin\theta\cos\phi, \sin\theta\sin\phi, \cos\theta\}.
\end{equation}
The photon transverse momentum is
\begin{equation}
    k_\perp = \omega\sin\theta.
\end{equation}
In the Coulomb gauge with $\bm k \cdot \bm e_{\lambda}(\bm{k}) = 0$, we have \cite{Knyazev2018Feb}

\begin{equation}
\begin{aligned}
& e^{\mu} = \{0, \bm e_{\lambda}\},\ \bm e_{\lambda}(\bm{k}) = \sum\limits_{\sigma = 0,\pm 1}\exp\{-i\sigma\phi\}d^1_{\sigma\lambda}(\theta)\bm \chi_{\sigma},\cr
& \bm \chi_0 = (0,0,1),\ \bm \chi_{\pm 1} = \mp\frac{1}{\sqrt{2}}(1,\pm i,0),
\end{aligned}
\end{equation}
where $d^1_{\sigma\lambda}(\theta)$ are the small Wigner functions\cite{Varshalovich} (Eq. \ref{eq:wigner}). Vectors $\bm \chi_{\sigma}$ represent eigenvectors for the photon spin operator $\hat{s}_z$
\begin{equation}
\hat{s}_z = \begin{pmatrix}
    0 & -i & 0\\
    i & 0 & 0\\
    0 &0 &0
\end{pmatrix},
\end{equation}

with the eigenvalues $\sigma = 0,\pm 1$,
\begin{equation}
\hat{s}_z \bm \chi_{\sigma} = \sigma \bm \chi_{\sigma}.    
\end{equation}
The decomposition of the polarization vector over the spin eigenvectors
$\bm\chi_\sigma$ explicitly separates the angular and polarization dependence
of the emitted photon. In this form, the polarization is encoded in the discrete
index $\sigma$, whereas the angular dependence is carried by the Wigner functions
$d^1_{\sigma\lambda}(\theta)$ and the azimuthal phase factor $e^{-i\sigma\phi}$.

\subsubsection{Initial and final states}

According to the solutions of transverse and longitudinal equations, we choose the initial and final states of the particle as $\Psi_i$ and $\Psi_f$. These states describe the particle before and after the photon emission, respectively. Since the emission does not occur in a definite point in space-time, we must not associate $\Psi_i$ and $\Psi_f$ with the states at the beginning and the end of acceleration. The nonstationary and nonlocal nature of the emission and acceleration processes is described by the matrix element \eqref{eq:Sfi_general}.

The final momentum $p_z'$ can also easily be misinterpreted as the kinetic one, and, therefore, it can be incorrectly assumed that it grows with time due to acceleration. However, in contrast to the kinetic one, the momentum $p_z$ and $p_z'$ are approximate quantum numbers and, as will be further seen, dictate the wave vector of the emitted photon $k_z$. We further discuss this meaning in Sec. \ref{sec:diff_prob} and Sec. \ref{sec:wide_packet}. 

\begin{figure}[h!]
    \centering
    \includegraphics[width=1\linewidth]{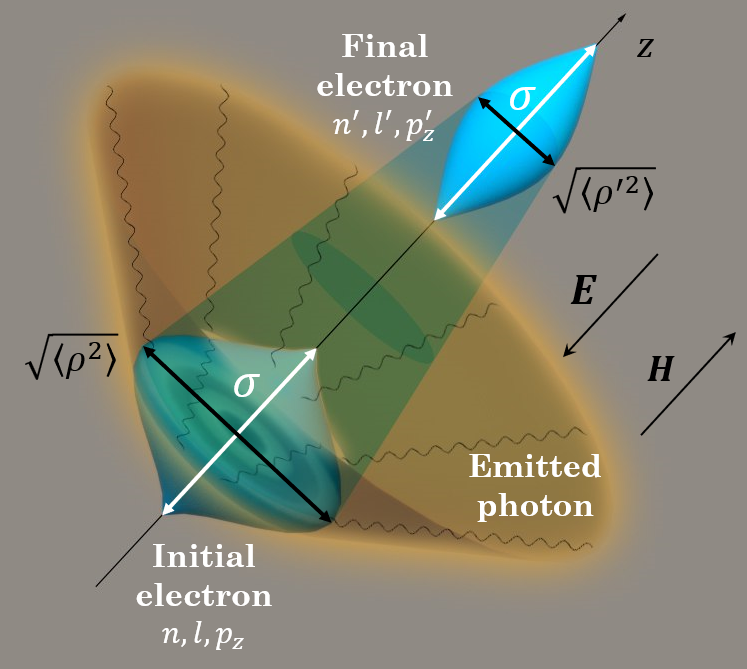}
    \caption{The electron is bounded by a magnetic field $\bm{H}$ and accelerated by an electric field $\bm{E}$, both of which are directed along the $z$-axis. During the acceleration the electron undergoes the transition between the Landau states with quantum numbers from $n,l,p_z$ to $n',l',p_z'$. The transition comes with the emission of the photon, thus, the root mean squared transverse radius of the Landau state changes from $\sqrt{\langle\rho^2\rangle} = \rho_{\rm H}\sqrt{2n + l + 1}$ to $\sqrt{\langle\rho'^2\rangle} = \rho_{\rm H}\sqrt{2n' + l' + 1}$ and becomes smaller. The longitudinal size of the electron packet, $\sigma$, remains unchanged during the emission.}
    \label{fig:illustration}
\end{figure}

The transition of the particle between the states with the quantum numbers $n, l$ and $n',l'$ comes with the emission of a photon, as illustrated in Fig. \ref{fig:illustration}. Thus, our initial and final states are

\begin{widetext}
\begin{equation}
\begin{aligned}
    &\Psi_i(\bm{r},t) = \frac{N}{\sqrt{\mathcal E_\text{cl}(t)}} \left( \frac{\rho}{\rho_{\text{H}}} \right)^{l} L_n^l \left( \frac{2 \rho^2}{\rho_{\text{H}}^2} \right) \exp{-\frac{\rho^2}{\rho_{\text{H}}^2} + i l \phi_r + ip_zz - \int\limits_0^t\mathcal E_\text{cl}(t')\dd t' -\frac{(z - z_{\text{cl}}(t) )^2}{2 \sigma^2}},\\
    &\Psi_f(\bm{r},t) = \frac{N'}{\sqrt{\mathcal E_\text{cl}'(t)}} \left( \frac{\rho}{\rho_{\text{H}}} \right)^{l'} L_{n'}^{l'} \left( \frac{2 \rho^2}{\rho_{\text{H}}^2} \right) \exp{-\frac{\rho^2}{\rho_{\text{H}}^2} + i l' \phi_r + ip_z'z - \int\limits_0^t\mathcal E_\text{cl}'(t')\dd t' -\frac{(z - z_{\text{cl}}'(t) )^2}{2 \sigma^2}},
\end{aligned} \label{eq:states}
\end{equation}
\end{widetext}
with the normalization constants
\begin{equation}
\begin{aligned}
    & N = \sqrt{\frac{2^{l^+1}}{\pi\rho_\text{H}^2}\frac{n!}{(n+l)!}} \frac{1}{\sqrt{(\pi\sigma^2)^{1/2}}}, \\
    & N' = \sqrt{\frac{2^{l^{'}+1}}{\pi\rho_\text{H}^2}\frac{n^{'}!}{(n^{'}+l^{'})!}} \frac{1}{\sqrt{(\pi\sigma^2)^{1/2}}}.
\end{aligned}
\end{equation}
Here $\mathcal E_\text{cl}(t)$ and $z_{\text{cl}}(t)$ are the classical energy and longitudinal trajectory of the
particle in the external fields, as introduced in Eqs. \ref{eq:state_longitudinal}. Their primed versions, $\mathcal{E}_\text{cl}'(t)$ and $z_\text{cl}'(t)$ contain $p_z'$ instead of $p_z$:
\begin{equation}
\begin{aligned}
    &\mathcal{E}_\text{cl}(t) = \sqrt{m^2+p_\perp^2 +(p_z+F_0t)^2},
    &z_{\text{cl}}(t) = \frac{\mathcal{E}_{\text{cl}}(t) - \mathcal{E}_{\text{cl}}(0) }{F_0},\\
    &\mathcal{E}'_\text{cl}(t) = \sqrt{m^2++ p_\perp'^2+(p_z'+F_0t)^2},
    &z'_{\text{cl}}(t) = \frac{\mathcal{E}'_{\text{cl}}(t) - \mathcal{E}_{\text{cl}}'(0) }{F_0}.
\end{aligned}
\end{equation}
The magnetic length $\rho_{\text{H}}$ is defined in
Eq.~\eqref{eq:magnetic_length}. 
The electron packet length $\sigma$ is the same in the initial and final
states and is not modified by the emission process. 
In general, emission of a photon can lead reduction of the electron coherence length, since the photon can partially resolve the internal structure of the wave packet \cite{Breuer2001, delisle2024decoherence}. A significant change of the coherence length, however, occurs when the emitted photon carries a noticeable fraction of the electron’s energy, so that its wavelength becomes comparable to the packet length. In this regime different longitudinal regions of the packet contribute to the emission with distinguishable phases, which leads to partial decoherence of the packet. Since the emission of photons with sufficiently high energies and short wavelengths is strongly suppressed, the change in the electron's coherence length can be safely neglected for the emission of soft X-rays and lower-frequency photons, whose wavelengths exceed the initial packet size. 

\subsubsection{Differential probability of emission}\label{sec:diff_prob}
For further calculations, the differential probability of emission needs to be obtained. The first-order (single vertex) probability in the scalar QED of a photon being emitted with the momentum between $\vb{k}$ and $\vb{k} + \dd\vb{k}$ is proportional to a squared module of the $S$-matrix element. 

The final longitudinal momentum of the particle lies within $p_z'$ and $p_z'+\dd p_z'$. Generally speaking, a set of states with different $p_z'$ is \textit{not} orthogonal due to the Gaussian factor \eqref{eq:alpha}. However, we can treat them as orthogonal if a proper measure of integration is defined. The integration of two states with different momenta over $z$ yields
\begin{equation}
\begin{aligned}
    \int\limits_{-\infty}^{+\infty}\Psi_{p_z'}^*\Psi_{p_z''} \dd z\propto \exp{-\frac{(z_\text{cl}'-z_\text{cl}'')^2}{4\sigma^2}-\frac{\sigma^2(p_z'-p_z'')^2}{4}}.
\end{aligned}
\end{equation}
When the energy gain $F_0t$ in the electric field significantly exceeds the longitudinal momentum difference $|p_z' - p_z''|$, the disparity between the two classical trajectories $z_\text{cl}'$ and $z_\text{cl}''$ becomes negligible. Consequently, the value of the integral depends mostly on the second term. We can express $|p_z'-p_z''|$ in terms of the emitted photon energies:
\begin{equation}
    |p_z' - p_z''| = |(p_z - p_z') - (p_z - p_z'')| = |\omega' - \omega''| \leqslant 2\omega.
\end{equation}In typical accelerator fields, the energy of a single photon is always
negligibly small compared to the total energy gain. Therefore, we treat two states as orthogonal if their momenta differ by at least $1/\sigma$. Thus, we arrive at the differential probability
\begin{equation}
    \dd W = \sum\limits_{\lambda= \pm1}|S_{fi}(\lambda)|^2\frac{V\dd^3\vb{k}}{(2\pi)^3} \sqrt{2}\sigma\dd p_z'. \label{eq:differential_probability}
\end{equation}
Here, the factor $\sqrt{2}\sigma$ is a measure associated with $\dd p_z'$. It is evaluated as a dispersion $\sigma_{p_z'}$ of the longitudinal momentum $p_z'$, calculated on a solution \eqref{eq:state_longitudinal}. For simplicity, we evaluate
\begin{equation}
\begin{aligned}
    \sigma_{z}^2 &\equiv  \langle z^2\rangle - \langle z \rangle^2 = \int \limits_{-\infty}^{+\infty}  j^0 z^2\dd z - \left[\int \limits_{-\infty}^{+\infty} j^0 z\dd z\right]^2 = \frac{\sigma^2}{2}.
\end{aligned}
\end{equation}
We further require the uncertainty relation to be minimized, obtaining
\begin{equation}
    \sigma_{p_z'}^2 = \frac{1}{2\sigma^2}. \label{eq:momentum_dispersion}
\end{equation}
With this choice, the set of packets labeled by $p_z'$ forms an
approximately orthonormal basis for the longitudinal motion, so that
Eq.~\eqref{eq:differential_probability} can be interpreted as the probability
density in the combined photon–electron phase space.

The detailed derivation of the $S$-matrix element is provided in the App. \ref{app:S_matrix}. Here we introduce the final result:

\begin{equation}
    \begin{aligned} \label{eq:Sfi_result}
    & S_{fi} = -2\pi i \frac{1}{\sqrt{2 \omega V}} \sqrt{\frac{2^{l+1}}{\pi\rho_\text{H}^2}\frac{n!}{(n+l)!}}\sqrt{\frac{2^{l'+1}}{\pi\rho_\text{H}^2}\frac{n'!}{(n'+l')!}} \\
    & \times \frac{e \rho_\text{H} }{2m}e^{i(l-l')\phi_k} \exp{-\frac{\sigma^2(p_z'-p_z+k_z)^2}{4}} \times\\
    & \times\sum\limits_{\sigma=0, \pm 1} i^{\sigma - l +l'}d^1_{\sigma,\lambda}(\theta)\mathcal T_\sigma(k_\perp \rho_\text{H}),
    \end{aligned}
\end{equation}
where the $\mathcal{T}$ functions are defined in Eq. \eqref{eq:time_integrals} and contain the only left integration over $t$. Eq.\eqref{eq:Sfi_result} clearly exhibits the main angular and kinematic
features of the process.
The
dependence on the azimuthal angle of the photon momentum, $\phi_k$, is fully
contained in the overall phase factor $e^{i(l-l')\phi_k}$, whereas the Wigner
functions $d^1_{\sigma\lambda}(\theta)$ encode the polar-angle
dependence. The sum over $\sigma=0,\pm1$ corresponds to different projections
of the photon spin; the functions $\mathcal T_\sigma(k_\perp\rho_\text{H})$ depend on the dimensionless
combination $k_\perp\rho_\text{H}$. The Gaussian factor $\exp{-\sigma^2(p_z' - p_z + k_z)^2/4}$ constrains the longitudinal momenta. It shows that conservation of the total $z$-momentum would require a completely delocalized state ($\sigma \to \infty$). However, since the acceleration process inherently involves spatial localization, strict momentum conservation is generally violated.
\subsection{Wide packet approximation} \label{sec:wide_packet}
Let us take a look at the damping factor in the $S$-matrix element \eqref{eq:Sfi_result}
\begin{equation}
    \exp{-\frac{\sigma^2(p_z'-p_z+k_z)^2}{4}} \label{sigma damping factor}.
\end{equation}
This factor suppresses the amplitude -- and therefore the probability -- exponentially everywhere except in a narrow region of momenta. It can be interpreted as a limitation either for the final electron momentum, or, equivalently, for the longitudinal photon momentum. The state of the emitted photon is beyond the scope of this work. Our main interest is the final electron state, especially its OAM. We therefore integrate over $k_z$ to obtain the total emission probability. The integral is localized near $k_z\simeq p_z-p'_z$.

The longitudinal length of the electron packet $\sigma$ is the main parameter determining the localization. Thus, for $\sigma\gg 1/|p'_z-p_z+k_z|$ the Gaussian can be approximated with a delta function. Equivalently, this requires that the longitudinal packet size in the laboratory frame exceeds the radiation formation length~\cite{BAIER2005261}:
\begin{equation}
    \sigma \gg L_\text{form}\sim \gamma^2\lambda_\text{ph},
\end{equation}
where $\lambda_\text{ph}$ is the photon wavelength and $\gamma= \mathcal{E}_\text{cl}(t)/m$ is the Lorentz factor. In this work, we do not consider formation zone effects, which would require a detailed analysis of the $\sigma \ll L_\text{form}$ regime. However, accounting for one would likely reduce the emission probability due to incomplete formation of radiation.
 
Now, let us treat \eqref{sigma damping factor} as a delta-like function. We will first take the squared module of an amplitude in order to avoid the problem of regularizing a square of a delta function.
\begin{equation}
    \abs{S_{fi}}^2 \propto \exp{-\frac{\sigma^2 (p_z'-p_z+k_z)^2} {2}} \sim \frac{\sqrt{2 \pi}}{\sigma} \delta(p_z'-p_z+k_z).
\end{equation}
The $\sqrt{2\pi}/\sigma$ factor is obtained from the delta function normalization condition by integrating both sides over $p_z'-p_z+k_z$ in infinite limits. From now on, we imply the relation $k_z = p_z - p_z'$ as a result of the integration, which allows us to simplify the integrals over $t$ in \eqref{eq:time_integrals}. The further integration over $t$, $k_\perp$ and $p_z'$ is numerical.

\section{Results}

\subsection{Suppression of large photon transverse momenta and its OAM}
\begin{figure}[h!]
    \centering
    \includegraphics[width=1\linewidth]{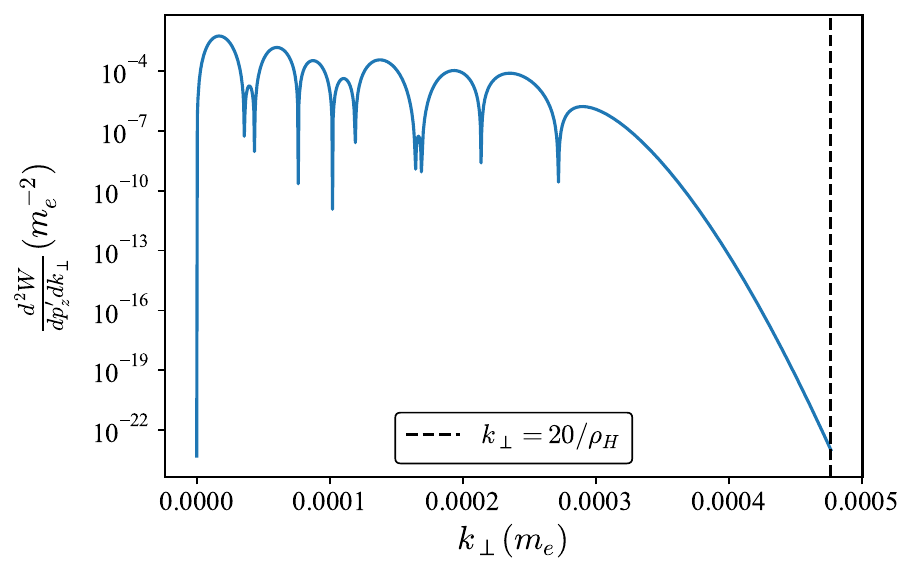}
    \caption{Double-differential probability, $d^{2}W/(dp'_z\,dk_\perp)$, of plane-wave photon emission by an accelerated vortex particle as a function of the photon transverse momentum $k_\perp$. The S-matrix is strongly localized around $k_{\perp\text{max}}\sim 2/\rho_\text{H}$. Parameters: $H=10~\mathrm{T}$, $E=10~\mathrm{MV/m}$, $\sigma=1~\mathrm{nm}$ (coherence length), $p_z=0.5~\mathrm{keV}$ (initial longitudinal electron momentum), $k_z= 3~\mathrm{eV}$ (longitudinal photon momentum), $n=n'=3$, $l = l' = 3$, $t_{\mathrm{out}}=25~\mathrm{ns}$ $(L= c t_\text{out} \approx 7~\mathrm{m})$.}

    \label{fig:double_diff_prob}    
\end{figure}

In order to obtain the lifetime of a state in the electromagnetic field, one needs to integrate over $k_\perp\in[0,\infty)$. However, due to the exponential factor in Eq.~\eqref{eq:mathcal_I} the integral can be taken in finite limits with the upper limit as following
\begin{equation}
    k_{\perp\text{max}} \sim \frac{2}{\rho_\text{H}} = m \sqrt{\frac{H}{H_c}}.
\end{equation}
This quantity defines the cutoff for $k_\perp$, the larger values of which turn out to be suppressed by the exponential factor, as can be seen for example in Fig.~\ref{fig:double_diff_prob}. The numerical results show that the electric field does not significantly modify this cutoff.



\begin{figure}[h]
    \centering
    \includegraphics[width=1
    \linewidth]{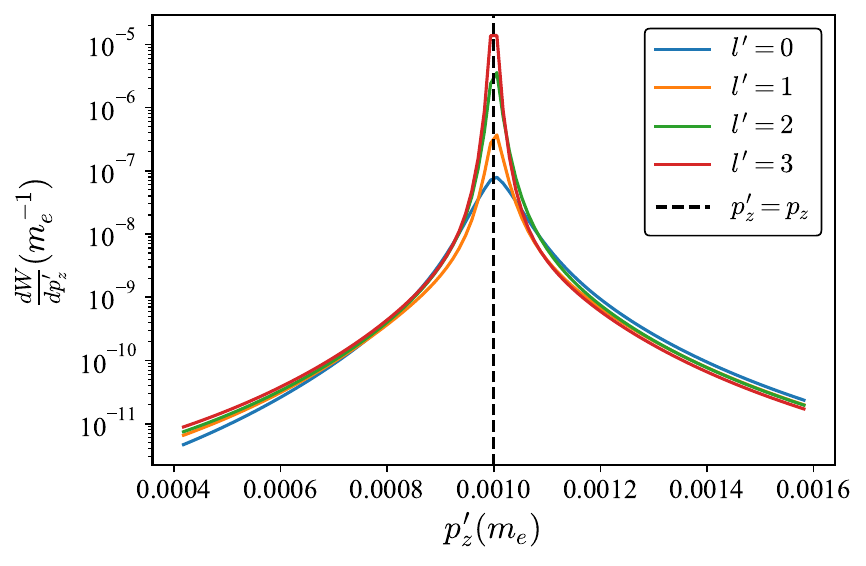}
    \caption{Differential probability of photon emission, $dW/dp'_z$, versus the final longitudinal momentum $p'_z$ for several final OAM projections $l'$ at fixed initial $l=3$. The distribution is peaked near $p'_z=p_z$, i.e., emission is dominated by soft photons $k_z=p_z-p'_z$. Parameters: $H=10~\mathrm{T}$, $E=10~\mathrm{MV/m}$, $\sigma=1~\mathrm{nm}$ (coherence length), $p_z=0.5~\mathrm{keV}$ (initial longitudinal electron momentum), $n=n'=3$, $t_{\mathrm{out}}=25~\mathrm{ns}$ $(L=c t_\text{out} \approx7~\mathrm{m})$.}
    \label{fig:dpz_prob_1e-3}
\end{figure}

The differential probability $\dd W/\dd p_z'$ (Fig. \ref{fig:dpz_prob_1e-3}) shows the principal dependence on $p_z'$ and  we see that it is localized around $p_z' = p_z$ and decreases as we move further from the localization point. 

\noindent This shows that the main contribution in photon emission comes from an electron with final states' longitudinal momentum close to the initial one. Equivalently, we can say that the emission occurs mainly due to low-energy photons, since $k_z = p_z - p_z'$. Therefore, integration over $p_z'$ comes with a narrow interval of $\pm 3\sigma_{p_z'}$ as follows
\begin{equation}
    p_z' \in \left[p_z - \frac{3}{\sqrt{2}\sigma}, p_z + \frac{3}{\sqrt{2}\sigma}\right],
\end{equation}
which coincides with the localization scale of the differential probability $\dd W/ \dd p_z'$. 

In addition, the OAM dependence is also predictable. As seen from Fig. \ref{fig:dpz_prob_1e-3}, it is much more possible to emit a photon with a \textit{small} OAM compared to the large ones. This shows that processes with large OAM losses are \textit{highly suppressed}.

\subsection{Emission probability per unit of time}

To characterize the scattering rate, we introduce the probability of photon emission per unit time,
\begin{equation}
    \dot W \equiv \frac{W}{t_\text{out}} .
\end{equation}
In stationary problems with the infinite process time, $t_\text{in} \to -\infty$, $t_\text{out} \to +\infty$, this comes as a natural way to get rid of non-physical infinite normalization time. In our case, $t_\text{out}$ is a physical time that we associate with the
duration of the particle acceleration. It is also convenient to consider the \textit{lifetime of the vortex state} $\tau$ instead of the transition rate: 
\begin{equation}
    \tau \equiv \dot{W}^{-1} = \frac{t_\text{out}}{W}.
\end{equation}
Note that the lifetime $\tau$, as we define it, depends on  $t_\text{out}$ in both terms: $W$ and $t_\text{out}$ itself. We interpret the lifetime $\tau$ as the characteristic time over which the particle preserves its OAM in the accelerator against radiative transitions. Thus, we denote the total probability of transitions with $n = n'$ from the state with OAM $l$ to the state with $l' \neq l$ as $W$. Transitions with $\Delta n\neq 0$ are strongly suppressed, similar to the case without the electric field \cite{Karlovets2023}, so the dominant contribution to $\tau$ comes from OAM-changing channels with $n=n'$.

\begin{widetext}

\begin{figure}[h]
    \centering
    \includegraphics[width=1\linewidth]{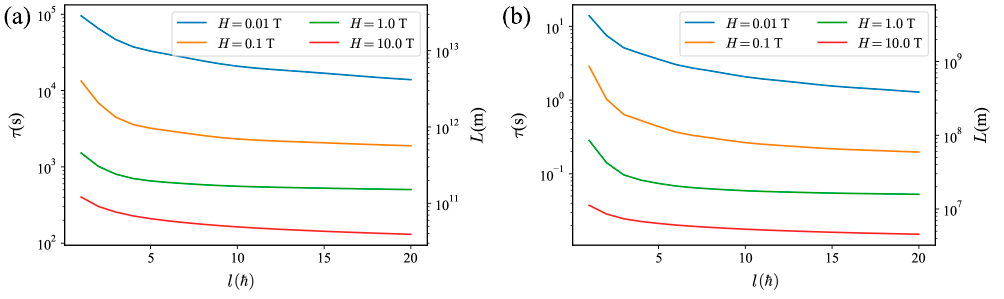}
    \caption{Lifetime $\tau$ of a vortex state against OAM-changing transitions as a
function of the initial OAM $l$ for different magnetic fields $H$. Decreasing
$H$ suppresses transverse emission and increases $\tau$.
(a) Parameters: $E=10~\mathrm{MV/m}$, $\sigma=1~\mathrm{nm}$ (coherence length), $p_z=0.5~\mathrm{keV}$ (initial longitudinal momentum), $n=n'=3$,
$t_{\mathrm{out}}=25~\mathrm{ns}$ $(L=c t_\text{out} \approx7~\mathrm{m})$.
(b) Same dependence for a higher energy and stronger field:
$E=100~\mathrm{MV/m}$, $p_z=100~\mathrm{keV}$,
$n=n'=10$, $t_{\mathrm{out}}=55~\mathrm{ns}$ $(L=c t_\text{out} \approx15~\mathrm{m})$;
the qualitative trend with $H$ remains the same.}

\label{fig:lifetime_mag}
\end{figure}
\end{widetext}

The dependence of the particle's lifetime on the initial OAM is shown in Fig. \ref{fig:lifetime_mag}. Here we see that $\tau \gg t_\text{out}$ for any achievable magnetic field $H$ even in the presence of an electric field. It is also understandable why the lifetime increases with the decreasing magnetic field. Radiative transitions occur due to the transverse dynamics and the change in discrete quantum numbers $n$ and $l$. The rms-radius of the ground Landau state scales as: 
\begin{equation}
    \rho_\text{H} = \sqrt{\frac{4}{|eH|}},\; \rho_\text{H}^2 \propto \frac{1}{H}.
\end{equation}
Therefore, as we lower the magnetic field, the probability decreases due to the factor $\exp{-(k_\perp\rho_\text{H})^2/8}$ in \eqref{eq:mathcal_I}, and the lifetime increases. 

\begin{widetext}

\begin{figure}[h]
        \centering
        \includegraphics[width=1\linewidth]{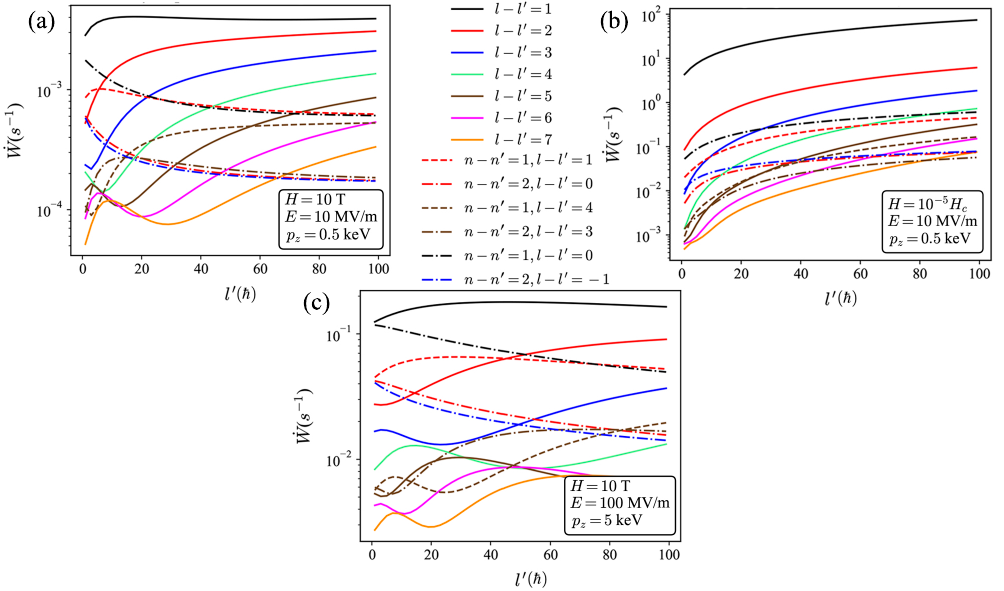}
        \caption{The total probability of photon emission per unit time $\dot{W}=W/t_{\mathrm{out}}$ for various scattering
channels as a function of the final OAM $l'$.
(a) Parameters: $H=10~\mathrm{T}$, $E=10~\mathrm{MV/m}$, $\sigma=1~\mathrm{nm}$ (coherence length), $p_z=0.5~\mathrm{keV}$ (initial longitudinal momentum), $n=n'=3$,
$t_{\mathrm{out}}=25~\mathrm{ns}$ $(L=c t_\text{out}\approx7~\mathrm{m})$.
The dominant contribution comes from channels with small OAM loss
$\Delta l=l-l'$, whereas transitions with $\Delta n\neq 0$ are suppressed.
(b) The same as in fig. (a) but for a stronger magnetic field,
$H=10^{-5}H_c$. The hierarchy of channels remains qualitatively the same,
with large $ \Delta l$ transitions being strongly suppressed.
(c) the same as in fig. (a), but for different packet width, acceleration time and quantum numbers: 
$H=10~\mathrm{T}$, $E=1~\mathrm{MV/m}$, $\sigma=1.75~\mathrm{nm}$,
$p_z=5~\mathrm{keV}$, $n=n'=20$, $t_{\mathrm{out}}=35~\mathrm{ns}$
$(L=c t_\text{out}\approx10~\mathrm{m})$. The qualitative dominance of small $\Delta l$
channels persists in this regime as well.}
        \label{fig:prob_per_unit_time_panel}
\end{figure}
\end{widetext}

To understand how the electric field affects different transitions with a change of OAM, let us consider probabilities per unit time $\dot{W}$ for various "scattering channels" (Fig.\ref{fig:prob_per_unit_time_panel}). Despite the complex channel structure, a clear trend emerges: transitions with a small OAM loss dominate over channels with a large $\Delta l$. For this set of parameters, an average value of the transition rate is $10^{-3} s^{-1}$. In statistical terms, only about one out of $10^3$ realizations per second leads to an OAM-changing emission. We also show graphs for stronger magnetic fields (Fig.\ref{fig:prob_per_unit_time_panel}, (b)) for illustrative purposes. We note that the probabilities in Fig. \ref{fig:prob_per_unit_time_panel} show the same qualitative behavior as those without the electric field.  

\begin{figure}[h]
    \centering
    \includegraphics[width=1\linewidth]{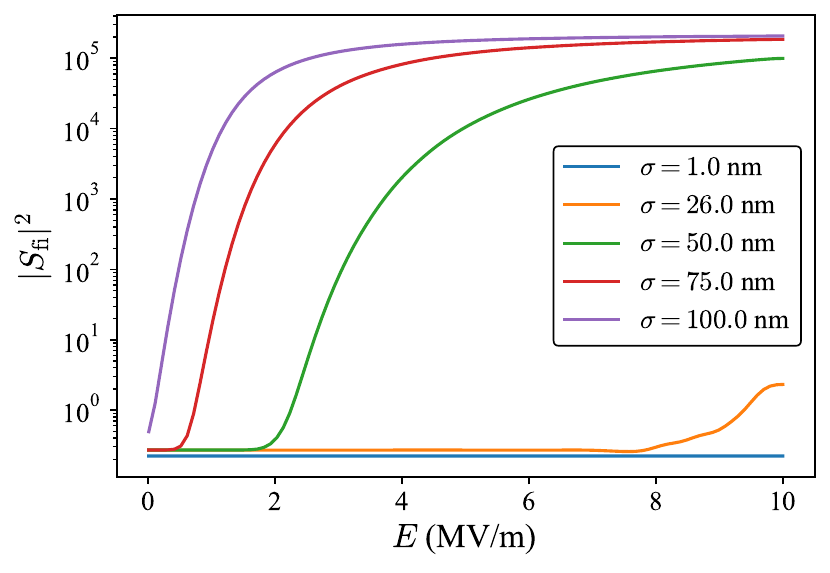}
    \caption{Squared S-matrix element $|S_{fi}|^{2}$ for the transition $n=n'=3$, $l=3\to l'=2$ as a function of accelerating electric field $E$ for different packet lengths $\sigma$. The graph shows the crossover from the narrow-packet regime (weak $E$-dependence) to the wide-packet regime (strong $E$-dependence). Parameters: $H=10~\mathrm{T}$, $p_z=0.5~\mathrm{keV}$ (initial longitudinal electron momentum), $k_z=3~\mathrm{eV}$ (longitudinal photon momentum), $t_{\mathrm{out}}=25~\mathrm{ns}$ $(L=7~\mathrm{m})$.}

    \label{fig:S_module_over_E}
\end{figure}

We further investigate the dependence on the electric field. Numerical modeling shows that the field significantly affects transition probabilities only within a narrow range of magnitudes, and that this range is controlled by the coherence length $\sigma$ (Fig.~\ref{fig:S_module_over_E}). Analytically, this is because $F_0$ and $\sigma$ enter Eqs. \eqref{eq:state_longitudinal} in a similar way, which determine the longitudinal state of the particle. 

In the parameter range relevant for accelerator facilities, the $S$-matrix, and hence the emission probability, is essentially \textit{independent of the field strength} for the realistic electron packet length, which for non-relativistic electrons is of the order of a few nanometers \cite{Ehberger, Cho2013, Karlovets2021Vortex},
\begin{equation}
\sigma\sim 1~\text{nm}.
\end{equation}
For relativistic electrons, this length is yet shorter than that due to the Lorentz contraction. At the same time for much wider packets, the probability exhibits a pronounced dependence on $E$.

The key factor is the radiation formation length, i.e. the longitudinal distance over which the emission process is coherently built up along the trajectory. If the coherence length of the packet in the laboratory frame, $\sigma$, is much smaller than the formation length, the packet is effectively point-like. In this regime, the detailed longitudinal profile of the packet \textit{no longer matters}: the radiation amplitude depends only on the instantaneous trajectory and becomes insensitive to how the charge density is distributed along $z$. Consequently, once $\sigma$ is reduced below the formation length, the emission probability saturates and is no longer changed by further variations of the external parameters. In the relativistic regime, the electron coherence length in the laboratory frame decreases as $\gamma^{-1}$ during acceleration, whereas the formation length of the photon $L_\text{form} \sim \gamma^{2}\lambda_\text{ph}$ grows much faster. As a result, the condition $L_\text{form} \gg \sigma$ is reached very quickly in practice for relativistic electrons, for which the radiation losses in a linac are tiny.

The electric field influences this picture mainly by modifying the classical trajectory and, therefore, the formation length itself. A stronger field increases the acceleration and curvature of the trajectory, and
thereby changes the longitudinal distance over which the radiation remains
phase-coherent. For initially long packets, in moderate fields the packet size is still comparable to or larger than the formation length over a substantial part of the motion; different longitudinal slices then radiate with different phases, and the probability is strongly affected by the value of $E$. As the field increases, the formation length becomes longer and the same packet enters the effectively point-like regime $\sigma \ll L_{\text{form}}$, so that the dependence on $E$ gradually disappears.

In contrast, an initially short packet, such as $\sigma \sim 1$ nm, is already much shorter than the relevant formation length even at relatively weak fields. Throughout the acceleration process, the radiation effectively “sees” a point-like source, because the coherence region always exceeds the packet size. In this situation, changing the electric field modifies the overall kinematics and spectral properties of the radiation, but does not resolve any additional longitudinal structure of the packet, and therefore the emission probability remains practically independent of the field magnitude.
\newpage

\subsection{Formation length}

The longitudinal part of the amplitude contains the overlap factor
\begin{equation}
    \exp{
        -\frac{\bigl(z_\text{cl}(t) - z'_\text{cl}(t)\bigr)^2}{4\sigma^2}
    },
    \label{eq:overlap_factor}
\end{equation}
where $z_\text{cl}(t)$ and $z'_\text{cl}(t)$ are the classical trajectories in the external field $F_0$ for two nearby longitudinal momenta. This factor directly controls how long along the trajectory the emission process remains coherent: once the separation 
$| z_\text{cl}(t) - z'_\text{cl}(t)|$ becomes larger than the intrinsic packet size $\sigma$, the overlap is exponentially suppressed and the corresponding contribution to the amplitude is negligible.

To make this more explicit, we consider two close initial momenta,
\begin{equation}
    p'_z = p_z + \delta p,
\end{equation}
and expand the primed trajectory to the first order in $\delta p$:
\begin{equation}
    z'_\text{cl}(t,p'_z)
    \simeq
    z_\text{cl}(t,p_z)
    +
    \frac{\partial z_\text{cl}(t)}{\partial p_z}\,\delta p.
\end{equation}
Using the explicit form of $z_\text{cl}(t)$ in a constant field $F_0$, we obtain
\begin{equation}
    \frac{\partial \mathcal E_\text{cl}(t)}{\partial p_z}
    =
    \frac{p_z + F_0 t}{\mathcal E_\text{cl}(t)}
    \equiv v(t),
\end{equation}
so that
\begin{equation}
    \frac{\partial z_\text{cl}(t)}{\partial p_z}
    =
    \frac{1}{F_0}\bigl[v(t)-v(0)\bigr].
\end{equation}
Therefore, the separation of the trajectories is
\begin{equation}
    z_\text{cl}(t) - z'_\text{cl}(t)
    \simeq
    -\,\frac{v(t)-v(0)}{F_0}\,\delta p.
    \label{eq:dz_approx}
\end{equation}

As we follow the longitudinal momentum conservation law, $k_z = - \delta p$. Therefore, the overlap factor \eqref{eq:overlap_factor} can then be written as
\begin{equation}
    \exp{-\frac{\Delta z(t)^2}{4\sigma^2}}
    \sim
    \exp{
        -\,\frac{\bigl(v(t)-v(0)\bigr)^2}{4 F_0^2 \sigma^2} k_z^2}.
    \label{eq:overlap_exp_estimate}
\end{equation}

We define the wave–packet–induced formation time $t_\text{form}$ as the moment when the exponent in \eqref{eq:overlap_exp_estimate} becomes of order unity,
\begin{equation}
    \frac{\bigl(v(t_\text{form})-v(0)\bigr)^2}{4 F_0^2 \sigma^2} k_z^2\sim 1,
\end{equation}
or equivalently
\begin{equation}
    | v(t_\text{form})-v(0)| \sim 2\,| F_0|\,\sigma /k_z.
    \label{eq:v_form_condition}
\end{equation}

Since $0 \le | v(t)| \le 1$, Eq.~\eqref{eq:v_form_condition} immediately leads to two regimes:

\paragraph{Narrow-packet regime.} 

If
\begin{equation}
    2\,| F_0|\,\sigma /k_z \ll 1,
    \label{eq:narrow_condition}
\end{equation}
then the condition \eqref{eq:v_form_condition} is met already at very small velocities, $| v(t_\text{form})-v(0)| \ll 1$. In this case, the coherence is lost at an early stage of acceleration, when the trajectory has not yet changed significantly. Thus, in the narrow-packet regime, the effective formation length is set purely by the wave–packet size and does not depend on the field strength.
This regime occurs for $\sigma \ll k_z/2|F_0|$.
\paragraph{Wide-packet regime.}

If
\begin{equation}
    2\,| F_0|\,\sigma/k_z \gtrsim 1,
    \label{eq:wide_condition}
\end{equation}
then the right-hand side of \eqref{eq:v_form_condition} is of order unity or larger, and the inequality $| v(t)| \le 1$ no longer restricts the overlap. In this case, the Gaussian factor \eqref{eq:overlap_factor} does not cut off the time integral at an early stage. The longitudinal structure of the packet remains visible over the relevant part of the trajectory, and the emission probability acquires a pronounced dependence on $F_0$.

It is convenient to characterize the transition between the two regimes by a critical packet length $\sigma_*(F_0)$ obtained from the equality in \eqref{eq:narrow_condition}–\eqref{eq:wide_condition}:
\begin{equation}
    \sigma_*(F_0) \sim \frac{k_z}{2| F_0|}.
    \label{eq:sigma_star_def}
\end{equation}

\paragraph{Numerical estimates}

To relate \eqref{eq:sigma_star_def} to the experimental parameters, we express $F_0$ in natural units. For an electric field of order $E = 10~\text{MV/m}$ one finds
\begin{equation}
    e E = 10~\text{MeV}/\text{m}.
\end{equation}
For $k_z = 3$ eV we have
\begin{equation}
    \sigma_*(F_0) \sim \frac{k_z}{2 eE} \sim \frac{3}{2 \cdot10^7} \; \text{nm}
    \sim 150 \; \text{nm}.
\end{equation}

In our simulations, narrow packets with $\sigma \sim 1~\text{nm}$ are well below $\sigma_*(F_0)$ for typical accelerator fields, so they fall into the regime \eqref{eq:narrow_condition}: the effective formation length is limited by the packet size, and the emission probability is practically independent of $F_0$. Wider packets start to become of the same order as $\sigma_*$ and belong to the regime \eqref{eq:wide_condition}: in this case the wave packet does not restrict the formation region, the field significantly modifies the classical trajectory over the relevant length, and the probability acquires a strong dependence on the electric field strength.

\begin{widetext}

\begin{figure}[h!]
    \centering
    \includegraphics[width=1\linewidth]{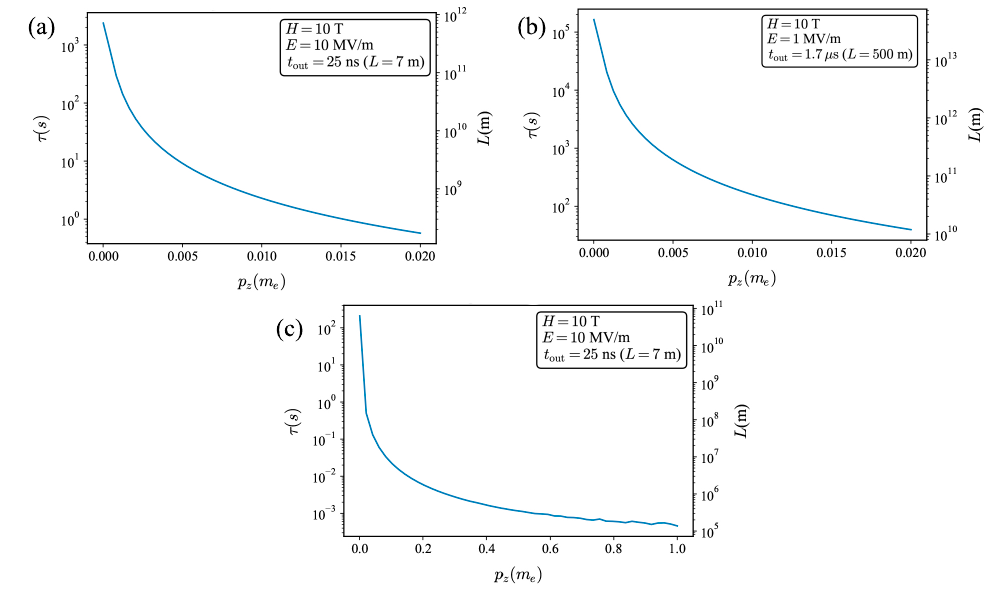}
    \caption{The lifetime $\tau$ versus the initial longitudinal momentum $p_z$ of the
accelerating vortex particle. Increasing $p_z$ (i.e., a larger Lorentz factor)
increases the probability of OAM-changing radiation and decreases the stability
of the vortex state. (a) Nonrelativistic momenta, parameters: $H=10~\mathrm{T}$, $E=10~\mathrm{MV/m}$,
$\sigma=1~\mathrm{nm}$ (packet length), $n=n'=3$, $l = 5$, $t_{\mathrm{out}}=25~\mathrm{ns}$
$(L=7~\mathrm{m})$.
(b) Nonrelativistic regime, same as (a), but for a longer acceleration section and lower electric field: $E=1~\mathrm{MV/m}$,
$t_{\mathrm{out}}=1.7~\mathrm{\mu s}$ $(L=500~\mathrm{m})$. (c) The same dependence as in  (a) but for the larger initial momenta.}

    \label{fig:lifetime_over_pzi_panel}
\end{figure}

\end{widetext}

Fig. \ref{fig:lifetime_over_pzi_panel} shows the dependence of the lifetime $\tau$ on the initial longitudinal momentum $p_z$. Increasing $p_z$ leads to a shorter lifetime. This trend is opposite to the naive classical expectation based on a time dilation, where one would anticipate $\tau'=\gamma\tau$ with $\gamma=\sqrt{m^2+p_\perp^2+(p_z+F_0 t)^2}/m$. The much steeper decrease observed in the nonrelativistic region highlights the genuinely quantum nature of the process, which is absent in a purely classical description.
\newpage


\section{Conclusion and discussion}
We have developed a nonstationary framework for photon emission by spinless charged particles with phase vortices accelerated in a linac. We have used the quasiclassical WKB method to obtain a localized wave packet in a longitudinal electric field, combined with Landau quantization in the transverse plane due to longitudinal magnetic field. This
avoids the pair-admixture issues of exact Klein–Gordon solutions and naturally incorporates a finite acceleration time and the radiation formation length.

Within this model, we have derived the first-order transition amplitude and emission probability, identified wide- and narrow-packet regimes, and showed that in the experimentally relevant narrow-packet regime the total emission probability is essentially insensitive to the electric field strength. The transverse magnetic dynamics predominantly governs OAM-changing transitions: channels with the large OAM loss are strongly suppressed, so the lifetimes of the vortex states {\it are extremely long compared to the acceleration time}, and they further increase for larger initial longitudinal momenta. 

Overall, we confirm the robustness of the vortex states in realistic accelerator fields and delineate the conditions under which the OAM-changing radiation can occur. Our formalism can be extended to Dirac particles, more complex accelerating structures, the spin–OAM coupling, and twisted-photon emission, and thus provides a basis for further studies of nonstationary radiation from the structured quantum states of charged particles.

Recent theoretical and simulation studies suggest that vortex beams can, in principle, retain their angular momentum in wakefields, especially in regimes where the beam propagates close to the axis in a symmetric or rotating wake~\cite{Vieira2018,Wu2019,WakeVortex2025}. These observations motivate the assumption that, on the scale of the of the wave packet, the wakefield can be approximated as locally homogeneous and weakly varying. However, this assumption is nontrivial: real wakefields exhibit strong spatiotemporal variations. Therefore, extending our conclusions to that regime would require a dedicated analysis that accounts for the full time-dependent structure of the wake. In particular, the radiative stability of vortex states in such complex backgrounds remains an open question that calls for future investigation.

\section{Acknowledgments}
This work is supported by the Russian Science Foundation under Grant No. 23-62-10026 \cite{RSF2023}.
We are grateful to I. Pavlov for fruitful discussions and assistance with the numerical code.

\appendix
\section{Transverse Klein-Gordon equation dynamics}
\subsection{Normalization constant}\label{app:transverse_norm}
The normalization constant for the transverse solution $\Psi_\perp$ is defined with the following condition
\begin{equation}
       \int\limits_0^\infty\rho\dd\rho\int\limits_0^{2\pi}\dd \phi_r|\Psi_\perp|^2 = 1.
\end{equation}
Using the associated Laguerre polynomial orthogonality property for integer $n,l$
\begin{equation}
    \int\limits_0^\infty \rho^l e^{-\rho} L_n^l(\rho) L_m^l(\rho) \dd \rho = \frac{(n+l)!}{n!} \delta_{n,m}
\end{equation}
We arrive at
\begin{equation}
    N_{\perp} = \sqrt{\frac{2^{l+1}}{\pi\rho_\text{H}^2}\frac{n!}{(n+l)!}}.
\end{equation}
\subsection{Transverse momentum}\label{app:transverse_momentum}
To check if the $C$ constant in Eq. \eqref{eq:KG_transverse} represents transverse momentum, we apply the corresponding $\hat{\vb{p}}_\perp$ operator to the $\Psi_\perp$ function in \eqref{eq:state_transverse}
\begin{equation}
\begin{aligned}
    &\hat{\vb{p}}_\perp^2 = (-i\nabla_\perp - e\vb{A}_\perp)^2,\\
    &\hat{\vb{p}}_\perp^2\Psi_\perp = \left(-\Delta_\perp +2ie\frac{A_{\phi_r}}{\rho}\partial_{\phi_r}+ e^2\vb{A}_\perp^2\right)\Psi_\perp.
\end{aligned}
\end{equation}
Eq. \eqref{eq:KG_transverse} can be used to simplify the derivation of $\Delta_\perp$. Taking into account that $\hat{l}_z = -i\partial_{\phi_r}$
\begin{equation}
\begin{aligned}
    &\hat{\vb{p}}_\perp^2\Psi_\perp = \left( eH\hat{l}_z+ C - \frac{e^2H^2\
    \rho^2}{4}-2e\frac{A_{\phi_r}}{\rho}\hat{l}_z +e^2 A_{\phi_r}^2\right)\Psi_\perp
\end{aligned}
\end{equation}
Since $A_{\phi_r} = H\rho/2$, all the terms except $C$ are canceled out and so 
$$
\hat{\vb{p}}_\perp^2\Psi_\perp = C \Psi_\perp.
$$

\section{The WKB approximation}
\subsection{Classical action derivation}\label{app:classical_action}
The classical action is a main part of the WKB approach, because it determines the final form of the solution. In order to evaluate it, the Hamilton-Jacobi problem has to be solved. Thus, let us consider the equation

\begin{equation}
    \pdv{S}{t} + H(\vb{r}, \grad S, t) = 0,
\end{equation}
with the Hamiltonian

\begin{equation}
    H(\vb{r},\vb{p},t) = \sqrt{m^2c^4+c^2\left(\vb{p}-\frac{e}{c}\vb{A}\right)^2} + e A^0.
\end{equation}

For clarity, in this section we consider a particle accelerating along the $z$ axis, with free transverse dynamics. The vector potential is
\begin{equation}
    A^\mu = \{0,Ect\,\bm{e}_z\} \quad \text{or} \quad A^\mu = \{Ez, 0\}.
\end{equation}
First, we solve the Hamilton's canonical equations to determine the classical path. As for the initial conditions, we set the coordinates and momentum of the particle at the beginning point $t_0=0$.

\begin{align}
\left\{ \begin{array}{l}
    \dot{\vb{r}} =  \displaystyle\pdv{H}{\vb{p}},\\
    \dot{\vb{p}} = - \displaystyle\pdv{H}{\vb{r}},\\
    \vb{r}(t_0) = \vb{r}_0,\\
    \vb{p}(t_0) = \vb{p}_0
\end{array} \right.
\end{align}
In both gauges, trajectories $\vb{r}$ and kinetic momentum $\vb{P}$ in the electric field are the same
\begin{equation}
\begin{cases}
    \vb{r}(t) \,\,= \vb{r_0} + \displaystyle\int\limits_{0}^tc^2\dfrac{\vb{P(t')}}{H(t')} \dd t',\\
    \vb{P}(t) = \vb{p}_0 - \dfrac{e}{c}\vb{A}(t)  \label{eq:solution_classical_trajectory}
\end{cases}
\end{equation}
However, the conjugate momentum depends on a gauge. In a nonstationary gauge $\vb{p} = \vb{p}_0$, whereas in a stationary one $\vb{p}(t) = F_0t \bm{e}_z$. Once we find the trajectories, the classical action at the initial point can be evaluated as follows \cite{BagrovMMf}
\begin{equation}
    S(\vb{r}_0,t) = S(\vb{r}_0, t_0) + \int \limits_{0}^t\left[ \vb{p(t',\vb{r_0})} \dot{\vb{r}}(t',\vb{r}_0) - H(t',\vb{r_0})\right]\dd t'
\end{equation}
The initial condition for the action is defined by the conjugate momentum at the starting point
\begin{equation}
    \vb{p}(t_0) = \nabla_{\vb{r}} S(\vb{r},t_0).
\end{equation}
This means that the action at the moment $t=t_0$ can be chosen as
\begin{equation}
    S(\vb{r}, t_0) = \vb{r}\cdot\vb{p}_0.
\end{equation}

To evaluate the action at any point in space and time, we need to substitute $\vb{r}_0 = \vb{r}_0 (\vb{r})$ from the previously calculated solution \eqref{eq:solution_classical_trajectory} into the function $S(\vb{r}_0,t)$. Thus, we get 
\begin{equation}
    S(\vb{r},t) = \vb{r}\cdot\vb{p}_0 - \int\limits_{0}^t\sqrt{m^2c^4+c^2p_{0\perp}^2+ c^2(p_{0z}+F_0t')}\dd t'
\end{equation}
for the time-dependent gauge, and
\begin{equation}
\begin{aligned}
    S(\vb{r},t) = \vb{r}\cdot\vb{p}_0 &- \int\limits_{0}^t\sqrt{m^2c^4+c^2p_{0\perp}^2+ c^2(p_{0z}+F_0t')}\dd t'\\
    &+F_0zt
\end{aligned}
\end{equation}
for the stationary one. In both expressions $p_{0\perp}^2 = p_{0x}^2+p_{0y}^2$. This difference causes a phase shift in a wavefunction, since the WKB solution we seek implies $\Psi_\parallel \propto \exp{i/\hbar \; S(\vb{r},t)}$. The transformation from one gauge to another is
\begin{equation}
\begin{aligned}
    &\vb{A} \to \vb{A} + \nabla f,\\
    &A^0 \to A^0 - \frac{1}{c}\frac{\partial}{\partial t}f,\\
    &f(\vb{r},t) = -Ec zt
\end{aligned}
\end{equation}
In the general Klein-Gordon theory, the phase factor appearing after a gauge transformation is
\begin{equation}
    \psi \to \psi \exp{i\frac{ef}{\hbar c}}
\end{equation}
which is exactly the phase shift caused by an action difference.
\subsection{The longitudinal solution}\label{app:longitudinal_solution}
The solution for problem \eqref{eq:longitudinal_wkb} can be found according to the following ansatz \cite{AkhiezerBerestetsky}:
\begin{equation}
    \Psi_\parallel (z,t) = \left(\sum\limits_{k=0}^\infty\hbar^ka_k(z,t)\right) \exp{\frac{i}{\hbar}S_\text{cl}(z,t)},
\end{equation}
where $a_k$ are the coefficients to be defined, $S_\text{cl}$ is a classical action, solution of a Hamilton-Jacobi problem for a classical particle in a homogeneous electric field. (App. \ref{app:classical_action}). Evaluating
\begin{equation}
\begin{aligned}
    &\left(i\hbar\partial_\mu-\frac{e}{c}(\mcA_{||})_\mu\right)\Psi_\parallel = \exp{\frac{i}{\hbar}S_\text{cl}}\\
    &\times\sum\limits_{k=0}^\infty \left(i\hbar^{k+1}(\partial_\mu a_k)  - \hbar^{k}a_k(\partial_\mu S_\text{cl}) - \frac{e}{c}(A_{||})_\mu \hbar^{k} a_k\right)
\end{aligned}
\end{equation}

We expand the left-hand side as a formal power series in $\hbar$. Setting each coefficient to zero yields an equation for each power of $\hbar$. Thus, the equation for $\hbar^0$ reads: 
\begin{equation}
    \left[\left(\partial^\mu S_\text{cl} + \frac{e}{c} (\mcA_{||})^\mu\right)^2-m^2c^2 - p_\perp^2\right]a_0 = 0.
\end{equation}
The $\hbar^0$ equation is automatically satisfied, since it represents the Hamilton-Jacobi equation itself. 

In the next $\hbar^1$ expansion order, we find
\begin{equation}
    \left[\partial^\mu\partial_\mu S_\text{cl} + 2 \left(\partial_\mu S_\text{cl} + \frac{e}{c}(\mcA_{||})_\mu\right)\partial^\mu\right]a_0 = 0. \label{eq: WKB a_0 correction}
\end{equation}
The classical kinetic momentum $P_\mu$ satisfies:
\begin{equation}
\begin{aligned}
    &P_\mu = - \left(\partial_\mu S_\text{cl}+ \frac{e}{c}( A_{||})_\mu\right),\\
    &P_\mu = m \frac{\dd r_\mu}{\dd \tau}, \; \; \tau = \gamma^{-1}t, \; \gamma = \frac{\mathcal{E}_\text{cl}}{mc^2},\\
    &\mathcal E_\text{cl} = \sqrt{m^2 + p_\perp^2 + (p_z+F_0t)^2}, \quad F_0 = - eE.
\end{aligned}\label{eq:classical_kinetic_momentum}
\end{equation}
Where $\tau$ is the time in the particle rest frame, $t$ is the laboratory time, and $\gamma$ is the Lorentz factor. The sign of the force $F_0$ is positive for the electron that accelerates along $z$ axis.

Taking this into account, Eq. \eqref{eq: WKB a_0 correction} can be rewritten as
\begin{equation}
    \left[\partial^\mu\partial_\mu S_\text{cl}-\frac{2\mathcal E_\text{cl}(t)}{c^2}\frac{\dd}{\dd t}\right]a_0 = 0.
\end{equation}
Since $\partial^\mu\partial_\mu S_\text{cl} = - \dot{\mathcal E}_\text{cl}/c^2$, the above equation has a solution of
\begin{equation}
    a_0(z,t) = \frac{\alpha(z,t)}{\sqrt{\mathcal E_\text{cl}(t)}}, \label{eq:a_0}
\end{equation}
where $\alpha$ is an arbitrary function satisfying
\begin{equation}
\begin{aligned} \label{eq: WKB alpha conditions}
    &\frac{\dd \alpha(z,t)}{\dd t} = 0,
    \quad \alpha(z_\text{cl},t) = \text{const},\\
    &\frac{\dd }{\dd t} = \frac{\partial }{\partial t} + \vb{v} \cdot \nabla.
\end{aligned}
\end{equation}
here $z_\text{cl} =\left( \mathcal{E}_{\text{cl}}(t) - \mathcal{E}_{\text{cl}}(0) \right)/F_0 $ is a classical trajectory \cite{LandauLifshitz2}.

The simplest way to satisfy the conditions \eqref{eq: WKB alpha conditions} is to choose $\alpha$ as $\alpha(z,t) = f(z-z_\text{cl}(t))$. Since we consider a particle acceleration process, it is natural to think of a wave packet, initially localized in the $z$-direction. Thus, we choose $\alpha(z,t)$ as follows:
\begin{equation}
    \alpha(z,t) = \exp{-\frac{(z-z_\text{cl}(t))^2}{2\sigma^2}}.\label{eq:alpha}
\end{equation}
Here, $\sigma$ is a length of the wave packet in the laboratory frame of reference. Hence, in the first order of the WKB approach an accelerated particle is localized around its classical trajectory \cite{Bagrov1993}.

\subsection{Accuracy of the solution}\label{app:accuracy}
The accuracy of the solution \eqref{eq:state_longitudinal} is determined similarly to the standard WKB procedure for the Schr\"{o}dinger equation. One can neglect the $\mathcal O(\hbar^2)$ terms when:
\begin{equation}
    \left|\frac{\hbar \partial_\mu\partial^\mu a_0}{a_0\partial_\mu\partial^\mu S_\text{cl} }\right| \ll 1,
\end{equation} where $a_0$ is the amplitude corresponding to $\hbar^0$ (see Eqs. \eqref{eq:a_0}-\eqref{eq: WKB alpha conditions}). This condition simplifies to
\begin{equation}
    \frac{\partial_\mu\partial^\mu a_0}{a_0} \ll \frac{\dot{\mathcal E_\text{cl}}}{\hbar c^2},
\end{equation}
Where dot stands for the laboratory time derivative. Applying exact form of $a_0(z,t)$ and supposing $z - z_\text{cl}(t) \sim \sigma$, one finds
\begin{equation}
    \frac{3}{4}\left(\frac{\dot{\mathcal E}_\text{cl}}{\mathcal E_\text{cl}}\right)^2 - \frac{1}{2}\frac{\ddot{\mathcal E}_\text{cl}}{\mathcal E_\text{cl}} + \frac{1}{F_0\sigma}\left(\ddot{\mathcal E}_\text{cl}- \frac{\left(\dot{\mathcal E}_\text{cl}\right)^2}{\mathcal E_\text{cl}}\right) \ll \frac{\dot{\mathcal E}_\text{cl}}{\hbar c^2}.
\end{equation}  
This inequality can be satisfied when:
\begin{equation}
    \frac{F_0\hbar c}{\mathcal E_\text{cl}^2} \ll 1, \quad \frac{F_0\hbar c^3(m^2c^2 + p_\perp^2)}{\mathcal E_\text{cl}^4} \ll 1, \quad \frac{\hbar c}{\sigma\mathcal E_\text{cl}} \ll 1.
\end{equation}
Each of the three inequalities corresponds to a
natural small parameter of the problem. The first two conditions,
\begin{equation}
    \frac{F_0\hbar c}{\mathcal E_\text{cl}^2} \ll 1,
    \qquad
    \frac{F_0\hbar c^3(m^2c^2 + p_\perp^2)}{\mathcal E_\text{cl}^4} \ll 1,
\end{equation}
quantify how weak the electric field is compared to the
characteristic QED (Schwinger) scale. For
electron energies not smaller than the rest energy,
$\mathcal E_\text{cl}\gtrsim m c^2$, one can estimate
\begin{equation}
    \frac{F_0\hbar c}{\mathcal E_\text{cl}^2}
    \sim \frac{eE_0\hbar}{m^2c^3}
    \lesssim \frac{E}{E_\text{c}},
\end{equation}
where $E_\text{c}=m^2c^3/(e\hbar)$ is the Schwinger critical field.

The second inequality contains an
extra factor $(m^2c^2 + p_\perp^2)/\mathcal E_\text{cl}^2 \leq 1$, so the
corresponding condition is automatically satisfied once the first one holds.
The third condition,
\begin{equation}
    \frac{\hbar c}{\sigma\,\mathcal E_\text{cl}} \ll 1,
\end{equation}
has a clear interpretation in terms of the de Broglie wavelength. The
quantity $\lambda_\text{dB} = 2\pi\hbar c/\mathcal E_\text{cl}$ is the
relativistic de Broglie wavelength of the electron, so this inequality can be
rewritten as
\begin{equation}
    \frac{\hbar c}{\sigma\,\mathcal E_\text{cl}}
    = \frac{\lambda_\text{dB}}{2\pi\sigma} \ll 1,
\end{equation}
i.e. the packet length $\sigma$ is \textit{much larger} than the de
Broglie wavelength. This is precisely the standard WKB requirement that the wave–packet envelope varies on a spatial scale much larger than the local
de Broglie wavelength.

For the parameter range treated in this work, all three dimensionless
parameters are many orders of magnitude smaller than unity. The typical
accelerating gradients of $E\sim 10$–$100~\mathrm{MV/m}$ are far below the
Schwinger field $E_\text{c}\simeq 1.32\times10^{16}~\mathrm{V/cm}$, so
$F_0\hbar c/\mathcal E_\text{cl}^2 \lesssim E/E_\text{c}\sim 10^{-11}$–$10^{-10}$,
and the second parameter is even smaller due to
$(m^2c^2 + p_\perp^2)/\mathcal E_\text{cl}^2 \leq 1$. For the smallest
energies used in our simulations, $\mathcal E_\text{cl}\simeq 511~\mathrm{keV}$
and $\sigma\sim 1~\mathrm{nm}$, we have
$\lambda_\text{dB} = 2\pi\hbar c/\mathcal E_\text{cl}\sim 10^{-2}~\mathrm{nm}$ and
$\lambda_\text{dB}/2\pi\sigma \sim 10^{-3}$.
At higher energies this ratio further decreases. Moreover, all three parameters introduced here are at most of order $\alpha \simeq 1/137$, and typically much smaller, i.e. below the magnitude of second–order QED corrections to the emission probability.  Thus, the WKB
approximation for the longitudinal wave function $\Psi_\parallel(z,t)$ is fully justified for all the parameters used.

\subsection{Relativistic normalization}\label{app:longinudinal_norm}
Since we solve a Klein-Gordon equation, the normalization condition will be connected with a conserving $j^0$ charge
\begin{equation}
   j^0 = \Im \left( \Psi_\parallel \left[ \frac{1}{c}\frac{\partial}{\partial t} - i \frac{e}{\hbar c} A^0 \right] \Psi^*_\parallel \right), \quad \int j^0 \dd V = 1.
\end{equation}
Evaluation of the charge in both gauge results in
\begin{equation}
    j^0 = \frac{|N_\parallel|^2}{c}\exp{-\frac{(z-z_\text{cl}(t))^2}{\sigma^2}}.
\end{equation}
Therefore, the longitudinal normalization constant is independent of the gauge and equals
\begin{equation}
    N_\parallel = \sqrt{\frac{c}{(\pi\sigma^2)^{1/2}}}.
\end{equation}

\section{$S$-matrix element calculation}\label{app:S_matrix}
We evaluate the $S$-matrix:
\begin{equation}
    S_{fi} = -ie\int d^4x\, j_{fi}^{\mu}(x)A_{\mu}^*(x),
\end{equation}
as it is defined in \eqref{eq:Sfi_general}. The transition current $j_{fi}^\mu(x)$ contains initial and final states $\Psi_i$, $\Psi_f$ as in \eqref{eq:states}.
The small Wigner $d$-functions are \cite{Varshalovich}
\begin{equation}
\begin{aligned}    
&d_{\lambda \lambda'}^{\;\,1}(\theta) =\frac{1}{2} \left(1+\lambda \lambda'\cos\theta\right),\\
&d_{11}^{\;\,1} = \cos^2(\theta/2),\, d_{1-1}^{\;\,1} = \sin^2(\theta/2),\cr
& d_{\lambda 0}^{\;1}(\theta) =-d_{0 \lambda}^{\;1}(\theta)=-\frac{\lambda}{\sqrt{2}} \sin\theta,
d_{00}^{\;1}(\theta)=\cos\theta,\\
& \sum\limits_{\sigma = 0,\pm 1} d_{\sigma_1\sigma}^1(\theta)\,d_{\sigma_2\sigma}^1(\theta) = \delta_{\sigma_1\sigma_2},
\end{aligned}\label{eq:wigner}
\end{equation}

Let us now recall the relation \cite{Karlovets2023}
\begin{equation}
    e_{\mu}^{*}(\partial^{\mu} + ieA^{\mu}) = \sum\limits_{\sigma = 0, \pm 1}d^1_{\sigma,\lambda}\exp(i\sigma(\phi - \phi_r))\hat{X}_{\sigma},
\end{equation}
where
\begin{equation}
\begin{aligned}
    &\hat{X}_\sigma = \left\{\hat{X}_{+1},\hat{X}_{-1},\hat{X}_0\right\}, \\
    &\hat{X}_{+1}=\frac{1}{\sqrt{2}}\left(\frac{i}{\rho}\partial_{\phi_r}-\partial_\rho-\frac{|e|H\rho}{2}\right),\\
    &\hat{X}_{-1} = \frac{1}{\sqrt{2}}\left(\frac{i}{\rho}\partial_{\phi_r}+\partial_\rho-\frac{|e|H\rho}{2}\right),\\
    &\hat{X}_0 = \partial_z+iF_0t.
\end{aligned}
\end{equation}
After expanding the transition current, taking into account $\partial_\rho L_n^l(\rho) = -L_{n-1}^{l+1}(\rho)$ for $n\geq1$, we arrive at
\begin{equation}
    S_{fi} = -\frac{ie}{2m}\sum\limits_{\sigma=0,\pm1}d^1_{\sigma,\lambda}\int\dd^4 xe^{i(\omega t-\vb{k}\cdot\vb{r})}e^{i\sigma(\phi-\phi_r)}\psi_f^*\psi_i X_\sigma,
\end{equation}
where $X_{\pm1}, X_0$ are defined as
\begin{equation}
\begin{aligned}
&X_{+1}=i\frac{\sqrt{2}}{\rho_\text{H}} \left[2\tilde{\rho} \left(\frac{L_{n-1}^{l + 1}}{L_{n}^{l}} - \frac{L_{n'-1}^{l'+1}}{L_{n'}^{l'}}\right) - \frac{l}{\tilde{\rho}} - 2 \tilde{\rho}\right], \\
&X_{-1}= - i\frac{\sqrt{2}}{\rho_\text{H}} \left[2\tilde{\rho} \left(\frac{L_{n-1}^{l + 1}}{L_{n}^{l}} - \frac{L_{n'-1}^{l'+1}}{L_{n'}^{l'}}\right) + \frac{l'}{\tilde{\rho}} + 2 \tilde{\rho} \right],\\
&X_0 = -p_z-p_z'-2F_0t+i\frac{z_\text{cl}(t)-z_\text{cl}'(t)}{\sigma^2}.
\end{aligned}\label{eq:X_functions}
\end{equation}
For $n = 0$ case $\partial_\rho L_0^l = 0$ and these terms are absent.
The integration over azimuthal angle $\phi_r$ involves the integral representation of Bessel $J_l$ functions
\begin{equation}
\begin{aligned}
&\int\limits_0^{2\pi}\frac{\dd\phi_r}{2\pi}e^{il \phi_r+ix\cos\phi_r} = i^l J_l(x),\\
&\int\limits_0^{2\pi}\frac{\dd\phi_r}{2\pi}e^{il \phi_r-ix\cos\phi_r} = i^{-l} J_l(x).\label{eq:Bessel}
\end{aligned}
\end{equation}
The transverse integrals over $\rho$ are
\begin{equation}
\begin{aligned}
    \mathcal I _\sigma(y) = \rho_\text{H} \int\limits_0^\infty &\dd \rho\rho^{l+l'+1}X_\sigma(\rho)L_n^l(2\rho^2)L_{n'}^{l'}(2\rho^2) \\
    &\times J_{l-l'-\sigma}(y\rho)e^{-2\rho^2}.
\end{aligned}
\end{equation}
The $\mathcal I_\sigma$ integrals are nontrivial and derived in other paper \cite{Karlovets2023}:
\begin{equation}
\begin{aligned}
    &\mathcal I _0(y) = \rho_\text{H} X_0\mathcal F_{n,n'}^{l,l'}(y),\\
    & \mathcal I _{+1}(y) = -i\sqrt{2}\left(    2\mathcal F_{n,n'}^{l,l'+1}(y)+(n+l)\mathcal F_{n,n'}^{l-1,l'}(y)\right),\\
    & \mathcal I _{-1}(y) = -i\sqrt{2}\left(    2\mathcal F_{n,n'}^{l+1,l'}(y)+(n'+l')\mathcal F_{n,n'}^{l,l'-1}(y)\right),\\
    &\mathcal F_{n,n'}^{l,l'}(y) = \frac{(n'+l')!}{n!}\frac{1}{2^{3(n-n')+2l-l'+2}}y^{2(n-n')+l-l'}\\
    & \qquad \quad \, \, \, \times L^{n-n'+l-l'}_{n'+l'}(y^2/8)L^{n-n'}_{n'}(y^2/8)e^{-y^2/8}.
\end{aligned}\label{eq:mathcal_I}
\end{equation}

After evaluating the Gaussian integral over $z$, we finally arrive at 
\begin{equation}
    \begin{aligned}
    & S_{fi} = -2\pi i \frac{1}{\sqrt{2 \omega V}} \sqrt{\frac{2^{l+1}}{\pi\rho_\text{H}^2}\frac{n!}{(n+l)!}}\sqrt{\frac{2^{l'+1}}{\pi\rho_\text{H}^2}\frac{n'!}{(n'+l')!}} \\
    & \times \frac{e \rho_\text{H} }{2m}e^{i(l-l')\phi_k} \exp{-\frac{\sigma^2(p_z'-p_z+k_z)^2}{4}} \times\\
    & \times\sum\limits_{\sigma=0, \pm 1} i^{\sigma - l +l'}d^1_{\sigma,\lambda}(\theta)\mathcal T_\sigma(k_\perp \rho_\text{H}),
    \end{aligned}
\end{equation}
where the $\mathcal{T}_\sigma$ functions are
\begin{equation}
\begin{aligned}
    &\mathcal T_0 = \rho_\text{H} \mathcal F_{n,n'}^{l,l'}(y)\int\limits_{0}^T \dd t\;\frac{X_0(t)}{{\sqrt{\mathcal{E}_\text{cl}\mathcal{E}_\text{cl}'}}}\exp{-\frac{(z_\text{cl}-z_\text{cl}')^2}{4\sigma^2}}\\
    &\times\exp{i\left(\omega t - \int \limits_{0}^t(\mathcal E_\text{cl}-\mathcal E_\text{cl}')\dd t'-\frac{(z_\text{cl}+z_\text{cl}')(p_z'-p_z+k_z)}{2}\right)}, \\
&\mathcal T_{\pm1} = \mathcal I_{\pm 1}(y) \int\limits_{0}^T \dd t\;\frac{1}{{\sqrt{\mathcal{E}_\text{cl}\mathcal{E}_\text{cl}'}}}\exp{-\frac{(z_\text{cl}-z_\text{cl}')^2}{4\sigma^2}} \\
&\times \exp{i\left(\omega t - \int \limits_{0}^t(\mathcal E_\text{cl}-\mathcal E_\text{cl}')\dd t'-\frac{(z_\text{cl}+z_\text{cl}')(p_z'-p_z+k_z)}{2}\right)}. \label{eq:time_integrals}
\end{aligned}
\end{equation}

\bibliographystyle{unsrt}
\bibliography{references}

@article{Knyazev2018Feb,
	author = {B.A. Knyazev and V.G. Serbo},
	title = {{Beams of photons with nonzero orbital angular momentum projection: new results}},
	journal = {Phys. Usp.},
	volume = {61},
	number = {5},
	pages = {449--479},
	year = {2018},
	month = Feb,
	publisher = {Physics-Uspekhi},
	doi = {10.3367/UFNe.2018.02.038306}
}

@article{Bliokh2017,
	author = {K.Y. Bliokh and I.P. Ivanov and G. Guzzinati and L. Clark and R. {Van Boxem} and A. B{\'e}ch{\'e} and R. Juchtmans and M.A. Alonso and P. Schattschneider and F. Nori and J. Verbeeck},
	doi = {https://doi.org/10.1016/j.physrep.2017.05.006},
	issn = {0370-1573},
	journal = {Physics Reports},
	note = {Theory and applications of free-electron vortex states},
	pages = {1-70},
	title = {Theory and applications of free-electron vortex states},
	url = {https://www.sciencedirect.com/science/article/pii/S0370157317301515},
	volume = {690},
	year = {2017}}

@book{LandauLifshitz2,
Author = {L. D. Landau and E. M. Lifshitz},
Publisher = {Butterworth-Heinemann, Burlington, Massachusetts},
Title = {The Classical Theory of Fields},
Volume = {2},
Year = {1980}}

@article{Cho2013,
  title = {Electron Beam Coherency Determined from Interferograms of Carbon Nanotubes},
  author = {Cho, B. and Oshima, C.},
  journal = {Bulletin of the Korean Chemical Society},
  volume = {34},
  issue = {3},
  pages = {892–898},
  year = {2013},
  doi = {10.5012/BKCS.2013.34.3.892}
}

@article{Ehberger,
  title = {Highly Coherent Electron Beam from a Laser-Triggered Tungsten Needle Tip},
  author = {Ehberger, Dominik and Hammer, Jakob and Eisele, Max and Kr\"uger, Michael and Noe, Jonathan and H\"ogele, Alexander and Hommelhoff, Peter},
  journal = {Phys. Rev. Lett.},
  volume = {114},
  issue = {22},
  pages = {227601},
  numpages = {5},
  year = {2015},
  month = {Jun},
  publisher = {American Physical Society},
  doi = {10.1103/PhysRevLett.114.227601},
  url = {https://link.aps.org/doi/10.1103/PhysRevLett.114.227601}
}

@Article{Karlovets2021Vortex,
  author    = {Dmitry Karlovets},
  journal   = {New Journal of Physics},
  title     = {Vortex particles in axially symmetric fields and applications of the quantum {Busch} theorem},
  year      = {2021},
  month     = {mar},
  number    = {3},
  pages     = {033048},
  volume    = {23},
  doi       = {10.1088/1367-2630/abeacc},
  publisher = {IOP Publishing},
  url       = {https://doi.org/10.1088/1367-2630/abeacc},
}

@book{BagrovGitman,
Author = {V. Bagrov, D. Gitman},
Publisher = {Berlin [a. o.] : de Gruyter},
Title = {The Dirac equation and its solutions},
Year = {2014}}

@article{Nikishov1969,
	author = {Nikishov, A. and Ritus, V.},
	title = {{Radiation spectrum of an electron moving in a constant electric field}},
	journal = {Sov. Phys. JETP.},
	volume = {29},
	number = {1093},
	pages = {42},
	year = {1969}
}

@book{AkhiezerBerestetsky,
Author = {A. I. Akhiezer and V. B. Berestetsky},
Publisher = {Nauka},
Title = {Quantum Electrodynamics},
Year = {1981}}

@book{Varshalovich,
    Author = {D. A. Varshalovich and A. N. Moskalev and V. K. Khersonskii},
    Publisher = {Nauka},
    Title = {Quantum Theory of Angular Momentum},
    Year = {1975}}

@article{Karlovets2023,
    Author = {D. V. Karlovets and A. Di Piazza},
    Title = {Emission of twisted photons by a scalar charged particle in a strong magnetic field},
    Journal = {Phys. Rev. D},
    Volume = {108},
    Issue = {6},
    Pages = {063007},
    Year = {2023},
    DOI = {10.1103/PhysRevD.108.063007}
}

@article{BAIER2005261,
title = {Concept of formation length in radiation theory},
journal = {Physics Reports},
volume = {409},
number = {5},
pages = {261-359},
year = {2005},
issn = {0370-1573},
doi = {https://doi.org/10.1016/j.physrep.2004.11.003},
author = {V.N. Baier and V.M. Katkov},
}

@book{Polytitsyn_diffraction,
author = {A. P. Potylitsyn and M. I. Ryazanov and M. N. Strikhanov and A. A. Tishchenko},
year = {2011},
month = {01},
pages = {},
title = {Diffraction Radiation from Relativistic Particles},
volume = {239},
isbn = {978-3-642-12512-6},
journal = {Springer Tracts in Modern Physics},
doi = {10.1007/978-3-642-12513-3}
}

@book{FradkinGitman_unstable_vacuum,
    author = "Fradkin, E. S. and Gitman, D. M. and Shvartsman, Sh. M.",
    title = "{Quantum electrodynamics with unstable vacuum}",
    year = "1991"
}

@article{Bagrov1993,
doi = {10.1088/0305-4470/26/22/038},
url = {https://doi.org/10.1088/0305-4470/26/22/038},
year = {1993},
month = {nov},
publisher = {},
volume = {26},
number = {22},
pages = {6431},
author = {V G Bagrov and V V Belov and A Yu Trifonov},
title = {Theory of spontaneous radiation by electrons in a trajectory-coherent approximation},
journal = {Journal of Physics A: Mathematical and General},
}

@article{Nikishov1969_pair_production,
    author = "Nikishov, A. I.",
    title = "{Pair production by a constant external field}",
    journal = "Zh. Eksp. Teor. Fiz.",
    volume = "57",
    pages = "1210--1216",
    year = "1969"
}

@misc{DKGrosmanPavlovArxiv2025,
      title={Angular momentum dynamics of vortex particles in accelerators}, 
      author={D. Karlovets and D. Grosman and I. Pavlov},
      year={2025},
      eprint={2507.08763},
      archivePrefix={arXiv},
      primaryClass={physics.acc-ph},
      howpublished= {\url{https://arxiv.org/abs/2507.08763}}, 
}

@misc{Zmaga2025,
      title={Radiation of "breathing" vortex electron packets in magnetic field}, 
      author={G. V. Zmaga and G. K. Sizykh and D. V. Grosman and Qi Meng and Liping Zou and Pengming Zhang and D. V. Karlovets},
      year={2025},
      eprint={2509.21195},
      archivePrefix={arXiv},
      primaryClass={quant-ph},
      howpublished = {\url{https://arxiv.org/abs/2509.21195}}, 
}

@article{Pavlov2024,
  title = {Emission of twisted photons by a Dirac electron in a strong magnetic field},
  author = {Pavlov, I. and Karlovets, D.},
  journal = {Phys. Rev. D},
  volume = {109},
  issue = {3},
  pages = {036017},
  numpages = {16},
  year = {2024},
  month = {Feb},
  publisher = {American Physical Society},
  doi = {10.1103/PhysRevD.109.036017},
  url = {https://link.aps.org/doi/10.1103/PhysRevD.109.036017}
}

@book{SokolovTernov,
    author = {Sokolov, A. A. and Ternov, I. M.},
    title = {Relativistic Electron},
    publisher = {Nauka},
    address = {Moscow},
    year = {1974},
}

@book{BagrovMMF,
    author = {Bagrov, V. G.  and Belov V. V. and Zadorozhny V. N. and Trifonov A. Yu.},
    title = {Methods of mathematical physics},
    publisher = {Izdatel'stvo Nauchno-Tekhnicheskoi Literatury},
    address = {Tomsk},
    year = {2002},
}

@misc{RSF2023,
  author       = {{Russian Science Foundation}},
  title        = {Project No.~23-62-10026},
  year         = {2023},
  howpublished = {\url{https://rscf.ru/en/project/23-62-10026/}}
}

@article{IVANOV2022,
title = {Promises and challenges of high-energy vortex states collisions},
journal = {Progress in Particle and Nuclear Physics},
volume = {127},
pages = {103987},
year = {2022},
issn = {0146-6410},
doi = {https://doi.org/10.1016/j.ppnp.2022.103987},
url = {https://www.sciencedirect.com/science/article/pii/S0146641022000461},
author = {Igor P. Ivanov},
}

@article{KarlovetsSerbo2020,
  title = {Effects of the transverse coherence length in relativistic collisions},
  author = {Karlovets, Dmitry V. and Serbo, Valeriy G.},
  journal = {Phys. Rev. D},
  volume = {101},
  issue = {7},
  pages = {076009},
  numpages = {18},
  year = {2020},
  month = {Apr},
  publisher = {American Physical Society},
  doi = {10.1103/PhysRevD.101.076009},
  url = {https://link.aps.org/doi/10.1103/PhysRevD.101.076009}
}

@article{Ivanov2020,
  title = {Doing Spin Physics with Unpolarized Particles},
  author = {Ivanov, Igor P. and Korchagin, Nikolai and Pimikov, Alexandr and Zhang, Pengming},
  journal = {Phys. Rev. Lett.},
  volume = {124},
  issue = {19},
  pages = {192001},
  numpages = {6},
  year = {2020},
  month = {May},
  publisher = {American Physical Society},
  doi = {10.1103/PhysRevLett.124.192001},
  url = {https://link.aps.org/doi/10.1103/PhysRevLett.124.192001}
}

@article{IvanovSuperkick2024,
  title = {Unambiguous Detection of High-Energy Vortex States via the Superkick Effect},
  author = {Li, Zhengjiang and Liu, Shiyu and Liu, Bei and Ji, Liangliang and Ivanov, Igor P.},
  journal = {Phys. Rev. Lett.},
  volume = {133},
  issue = {26},
  pages = {265001},
  numpages = {6},
  year = {2024},
  month = {Dec},
  publisher = {American Physical Society},
  doi = {10.1103/PhysRevLett.133.265001},
  url = {https://link.aps.org/doi/10.1103/PhysRevLett.133.265001}
}

@article{VortexMuon,
  title = {Decay of the vortex muon},
  author = {Zhao, Pengcheng and Ivanov, Igor P. and Zhang, Pengming},
  journal = {Phys. Rev. D},
  volume = {104},
  issue = {3},
  pages = {036003},
  numpages = {11},
  year = {2021},
  month = {Aug},
  publisher = {American Physical Society},
  doi = {10.1103/PhysRevD.104.036003},
  url = {https://link.aps.org/doi/10.1103/PhysRevD.104.036003}
}

@article{IvanovSuperkick2022,
  title = {Observability of the superkick effect within a quantum-field-theoretical approach},
  author = {Ivanov, Igor P. and Liu, Bei and Zhang, Pengming},
  journal = {Phys. Rev. A},
  volume = {105},
  issue = {1},
  pages = {013522},
  numpages = {12},
  year = {2022},
  month = {Jan},
  publisher = {American Physical Society},
  doi = {10.1103/PhysRevA.105.013522},
  url = {https://link.aps.org/doi/10.1103/PhysRevA.105.013522}
}

@article{Meng2025PRR,
  title = {Relativistic quantum mechanics of charged vortex particles accelerated in a uniform electric field},
  author = {Meng, Qi and Huang, Ziqiang and Liu, Xuan and Ma, Wei and Yang, Zhen and Lu, Liang and Silenko, Alexander J. and Zhang, Pengming and Zou, Liping},
  journal = {Phys. Rev. Res.},
  volume = {7},
  issue = {4},
  pages = {043213},
  numpages = {6},
  year = {2025},
  month = {Nov},
  publisher = {American Physical Society},
  doi = {10.1103/z6j7-grs2},
  url = {https://link.aps.org/doi/10.1103/z6j7-grs2}
}

@article{Breuer2001,
  title = {Destruction of quantum coherence through emission of bremsstrahlung},
  author = {Breuer, Heinz-Peter and Petruccione, Francesco},
  journal = {Phys. Rev. A},
  volume = {63},
  issue = {3},
  pages = {032102},
  numpages = {18},
  year = {2001},
  month = {Feb},
  publisher = {American Physical Society},
  doi = {10.1103/PhysRevA.63.032102},
  url = {https://link.aps.org/doi/10.1103/PhysRevA.63.032102}
}

@misc{delisle2024decoherence,
      title={Decoherence of a 2-Path System by Infrared Photons}, 
      author={Colby DeLisle and P. C. E. Stamp},
      year={2024},
      eprint={2211.05813},
      archivePrefix={arXiv},
      primaryClass={quant-ph},
      url={https://arxiv.org/abs/2211.05813}, 
}

@article{Vieira2018,
  title = {Optical Control of the Topology of Laser-Plasma Accelerators},
  author = {Vieira, J. and Mendon\ifmmode \mbox{\c{c}}\else \c{c}\fi{}a, J. T. and Qu\'er\'e, F.},
  journal = {Phys. Rev. Lett.},
  volume = {121},
  issue = {5},
  pages = {054801},
  numpages = {6},
  year = {2018},
  month = {Jul},
  publisher = {American Physical Society},
  doi = {10.1103/PhysRevLett.121.054801},
  url = {https://link.aps.org/doi/10.1103/PhysRevLett.121.054801}
}

@article{WakeVortex2025,
author = {Huang, Rong and Chen, Min and Geng, Panfei and Zhu, Xinzhe and An, Xiangyan and Sheng, Zhengming},
year = {2025},
month = {06},
pages = {},
title = {Wakefield generation and electron acceleration driven by a spatiotemporal vortex in plasma},
volume = {32},
journal = {Physics of Plasmas},
doi = {10.1063/5.0263070}
}

@article{Wu2019,
doi = {10.1088/1367-2630/ab2fd7},
url = {https://doi.org/10.1088/1367-2630/ab2fd7},
year = {2019},
month = {jul},
publisher = {IOP Publishing},
volume = {21},
number = {7},
pages = {073052},
author = {Wu, Yitong and Ji, Liangliang and Geng, Xuesong and Yu, Qin and Wang, Nengwen and Feng, Bo and Guo, Zhao and Wang, Weiqing and Qin, Chengyu and Yan, Xue and Zhang, Lingang and Thomas, Johannes and Hützen, Anna and Büscher, Markus and Rakitzis, T Peter and Pukhov, Alexander and Shen, Baifei and Li, Ruxin},
title = {Polarized electron-beam acceleration driven by vortex laser pulses},
journal = {New Journal of Physics},
}

@book{MaslovFedoruk,
  author    = {Maslov, V. P. and Fedoruk, M. V.},
  title     = {Semi-Classical Approximation in Quantum Mechanics},
  year      = {1981},
  publisher = {D. Reidel Publishing Company},
  address   = {Dordrecht},
  series    = {Mathematical Physics and Applied Mathematics},
  volume    = {7},
  note      = {Russian original: Maslov V.P., Fedoruk M.V. \emph{Kvaziklassicheskoe priblizhenie dlya uravnenij kvantovoj mekhaniki}. Moscow: Nauka, 1976.},
  language  = {english}
}

@misc{Dyatlov2025,
      title={High-OAM Deep Ultraviolet Twisted Light Generation for RF-Photoinjector Applications}, 
      author={A. S. Dyatlov and D. M. Dolgintsev and V. V. Gerasimov and V. V. Kobets and V. P. Nazmov and M. A. Nozdrin and A. N. Sergeev and D. S. Shokin and K. E. Yunenko and D. V. Karlovets},
      year={2025},
      eprint={2512.08442},
      archivePrefix={arXiv},
      primaryClass={quant-ph},
howpublished = {\url{https://arxiv.org/abs/2512.08442}},
      url={https://arxiv.org/abs/2512.08442}, 
}

\end{document}